\documentclass[%
 reprint,
 amsmath,amssymb,
 aps,
]{revtex4-2}

\usepackage{graphicx}
\usepackage{dcolumn}
\usepackage{bm}
\usepackage{stmaryrd}
\usepackage{bbm}
\usepackage{amssymb}
\usepackage{algorithm}
\usepackage{algorithmic}

\newcommand\X{\bm{X}}
\newcommand\x{\bm{x}}
\newcommand\Y{\bm{Y}}
\newcommand\K{\bm{K}}
\newcommand\W{\bm{W}}
\newcommand\Q{\bm{Q}}
\newcommand\q{\bm{q}}
\newcommand\eps{\bm{\epsilon}}
\newcommand\m{\bm{m}}

\newcommand\SSel{\mathcal{S}}
\newcommand\T{\bm{R}}
\newcommand\Tr{\mathrm{tr}}
\newcommand\Z{\mathcal{Z}}

\newcommand\sub{\mathcal{S}}

\begin{document}

\preprint{}

\title{Large Deviations of Semi-supervised Learning in the Stochastic Block Model}

\author{Hugo Cui}

\author{Luca Saglietti}%
\author{Lenka Zdeborov\'a}%

\affiliation{SPOC laboratory,Physics Department, \'Ecole Polytechnique F\'ed\'erale de Lausanne.
}%

\date{\today}

\begin{abstract}
In community detection on graphs, the semi-supervised learning problem entails inferring the ground-truth membership of each node in a graph, given the connectivity structure and a limited number of revealed node labels. Different subsets of revealed labels can in principle lead to higher or lower information gains and induce different reconstruction accuracies. In the framework of the dense stochastic block model, we employ statistical physics methods to derive a large deviation analysis for this problem, in the high-dimensional limit. This analysis allows the characterization of the fluctuations around the typical behaviour, capturing the effect of correlated label choices and yielding an estimate of their informativeness and their rareness among subsets of the same size. 
We find theoretical evidence of a non-monotonic relationship between reconstruction accuracy and the free energy associated to the posterior measure of the inference problem. We further discuss possible implications for active learning applications in community detection. 
\end{abstract}

\maketitle


\section{\label{sec:intro}Introduction}

Many problems in data science require learning from graph data, where relational as well as local information might be available. Leveraging on the graph structure is often sufficient for inferring a plausible partition of the nodes into a finite number of communities \cite{Kumaran2016CommunityFB,Abbe2017CommunityDA,Traud2011SocialSO}, each characterized by different internal and external connectivities. In some cases, the community reconstruction accuracy can be further improved by accessing the labels (i.e., ground truth community assignments) of a selected subset of nodes.
This semi-supervised learning setting is especially relevant when the hidden labels are expensive and time-consuming to obtain, which is often the case in real world applications \cite{Ping2018BatchMA,Liu2020ActiveLF,Gadde2016ActiveLF}. 

Inference processes on graphs are intrinsically based on a modeling assumption for the graph generation process. The stochastic block model (SBM) can arguably be seen as the canonical probabilistic model for graphs displaying a community structure \cite{Abbe2017CommunityDA,Decelle2011AsymptoticAO,Yan2014ModelSF}, able to capture the typical features of both assortative and disassortative graphs. In an instance of an SBM, with $N$ nodes and $r$ communities, one assumes the existence of a planted community assignment and that each edge in the graph is independently added with a probability that is function solely of the hidden labels of the involved nodes. Depending on the scaling of the edge probability with $N$, the SBM framework can be employed for modeling both dense and sparse graphs. In this work, we focus our analysis on the dense case, which simplifies the statistical physics description, in the high dimensional limit $N\to\infty$, by means of the replica method \cite{MezardMontanari}. In this setting, the theoretically interesting regime, leading to non-trivial inference, is one where the graph connectivity only retains a weak $\mathcal{O}(1/\sqrt{N})$ dependence on group membership \cite{review}.

The semi-supervised setting in the SBM was previously considered in \cite{Eaton2012ASM, ver2014phase, Zhang2014PhaseTI}. A crucial assumption in these works, however, is that the fraction of disclosed labels $\{x_i\}_{i\in\sub}$, where $\sub$ is a revealed subset of cardinal $|\sub|=nN$ and $0<n<1$, is chosen uniformly at random among all nodes. While this assumption is the simplest one from a theoretical point of view, in realistic settings the data collection process is often far from random. In fact, a wide variety of techniques ranging from centrality-based heuristics to more refined and expensive active learning strategies \cite{Ping2018BatchMA,Liu2020ActiveLF,Gadde2016ActiveLF}, could be employed in order to select node labels that are less redundant and carry more information than randomly sampled ones. 

In the present work, we depart from the typical (random) choice of the subset and we study the large deviation properties of semi-supervised learning in the SBM model. The theoretical goal is to characterize how correlations in the choice of the labelled nodes can impact the associated inference performance. In order to do this, we compare all possible revealed subsets of a given size $n$ and organize them in a statistical mechanical ensemble, where the associated free energy (i.e., log posterior measure) plays the role of an effective energy function \cite{obuchi2018statistical}. We can then obtain an estimate of the subset-entropy, i.e., the exponential rate of the number of subsets that achieve a given value of the free energy. This subset-entropy can be regarded as a characterization of the rareness of atypical subsets.
We then discuss the relationship between free energy and reconstruction accuracy.       

From the point of view of the replica method, studying the large deviations entails a computation akin to a 1RSB in structure \cite{Replica, Cui2020LargeDF}: the free entropy of the model and the entropy of labelled subsets are linked by a Legendre transform controlled by a temperature ($\beta$ in the following). In order to simplify the calculations, we restrict our analysis to the Bayes-optimal inference setting, where the parameters of the generative model are known and matched in the inference model. In fact, in this setting the Nishimori identities \cite{nishimori1980exact, Nishimori} guarantee replica symmetry to hold at least for the equilibrium solutions of the system. In the out-of-equilibrium settings spanned by the large deviation, however, the Nishimori conditions can be violated. We thus provide a stability analysis, checking for the appearance of possible replica symmetry breaking phenomena \cite{RSB, MezardVirasoro}. The calculations are presented for a more general class of planted inference problems (containing the SBM as a special case), namely symmetric low-rank matrix factorization problems \cite{Thibault_MMSE}. This more general framework can be useful for a direct connection with different machine learning applications, ranging from PCA to high-dimensional Gaussian mixture clustering \cite{Thibault_Constrained}.

{The main results we obtain can be summarized as follows:}
\begin{itemize}
    \item Using the replica method, we compute the entropy of ways to disclose a fraction $n$ of nodes yielding a given value of the free energy. We provide results for the case of a dense SBM with two balanced communities \cite{Thibault_MMSE,Thibault_Constrained}.
    \item We provide evidence of a non-monotonic dependence between the free energy and the reconstruction accuracy for high values of SNR, which we found somewhat unexpected. 
    \item We discuss the consequences for active community detection where one is free to choose the subset of nodes for which labels will be revealed. In particular, we unveil that maximizing the free energy is not always the best criterion to select subsets leading to maximal reconstruction accuracy. We illustrate this point by implementing Monte-Carlo selection strategies based on the Thouless-Anderson-Palmer (TAP) \cite{TAP} evaluation of the free energy and discussing their performance.
\end{itemize}

The remainder of the paper is organized as follows: in section \ref{sec:setting}, we formally introduce the models analyzed in this work and define the large deviation framework. In section \ref{sec:replica} we outline the replica computation of the subset-entropy, which is the main result of this paper. In section \ref{sec:results}, we show the predictions obtained from the large deviation analysis, describing the interplay between subset-entropy, free-energy and inference performance. In section \ref{sec:stability}, we provide some derivation details and the results of the stability analysis associated to the replica symmetric ansatz. Finally, in section \ref{sec:consequences}, we discuss possible implications of our results for active learning in community detection, comparing our predictions with MCMC simulations in finite size graphs.

\section{Setting} \label{sec:setting}
\subsection{Community detection in the SBM}
\paragraph{The inference problem} The stochastic block model on $N$ individuals and $r$ communities is a commonly used generative probabilistic model for graphs displaying community structure. In the SBM, each individual is first assigned independently at random to a community $x_i \in \llbracket 1,r\rrbracket$ according to a prior $P_X(\cdot)$. Between each pair of individual $(i,j)$, an edge is then created with probability $P_{\mathrm{out}}(y_{ij}=1|x_i,x_j)$, where $y_{ij}=\mathbbm{1}_{(ij)\in E}$ is the $i,j$ component of the adjacency matrix of the resulting graph. In the \textit{community detection} problem on the obtained graph, the statistician aims at inferring back the community of each individual $\bm{X}=(x_i)_{1\le i\le N}$ from the sole knowledge of the graph structure, or, equivalently, its adjacency matrix $\bm{Y}$.\\

\paragraph{SBM with two balanced communities} 
While the following derivation is detailed for generic SBM, we choose to illustrate our results for a dense SBM with two balanced communities of respective average sizes $N\rho$ and $N(1-\rho)$ and same average connectivity \cite{Thibault_Constrained,Thibault_MMSE}. More precisely, the dense SBM with two balanced communities is specified, for any $\rho,p\in(0,1),~\mu\ge0$, by the prior 
\begin{align}
    P_{X}(\cdot)=\rho\delta(\cdot-x_+)+(1-\rho)\delta(\cdot-x_-),
\label{eq:SBM_prior}
\end{align}
and the posterior
\begin{align}
    P_{\mathrm{out}}(y|x,x')=\delta_{y,1}\left(p+\mu \frac{xx'}{\sqrt{N}}\right)+\delta_{y,0}\left(1-p-\mu \frac{xx'}{\sqrt{N}}\right),
\label{eq:SBM_posterior}
\end{align}
with the shorthands $x_+:=\sqrt{(1-\rho)/\rho},~x_-=-1/x_+$ labelling the two communities. Therefore, two individuals pertaining to communities $x\in\{x_+,x_-\}$ and $x'\in\{x_+,x_-\}$ are connected with probability $p+\mu x x'/\sqrt{N}$, thereby defining a graph where edges link connected individuals. Notice that, in order to obtain an inference problem where the performance is neither as bad as random guessing nor perfect the difference between inter and intra-community connectivities needs to be small, of order $\mathcal{O}(1/\sqrt{N})$. The special case of an SBM with two balanced communities makes for a non-trivial inference problem, as the prior \eqref{eq:SBM_prior} and posterior \eqref{eq:SBM_posterior} imply that nodes in both communities have the same connectivity, and that therefore no information on the community can be easily accessed based on the sole degree of a node.

For the standard SBM problem with two balanced communities, \cite{Thibault_Constrained,Thibault_MMSE} evidenced the presence of a phase transition for some value of $\Delta= p(1-p)/\mu^2$ depending on $\rho$, above which no reconstruction is information-theoretically possible. This transition is of first order if $\rho\in (0,\frac{1}{2}-\frac{1}{\sqrt{12}})\cup(\frac{1}{2}+\frac{1}{\sqrt{12}},1)$, and is of second order otherwise. In the semi-supervised setting where a non-zero fraction of nodes $n>0$ is disclosed \cite{Zhang2014PhaseTI}, instead, partial community reconstruction is always achievable, the second order phase transition is smeared out, and first order phase transition shrinks and disappears in the tri-critical point as the number of revealed nodes increases.\\

\paragraph{Symmetric matrix factorization} An SBM can be recast more generally into a \textit{low rank symmetric matrix factorization problem} \cite{Thibault_Constrained,Thibault_MMSE}. In what follows, this formalism shall be employed for the sake of generality, while results will be given and discussed for the special case of the SBM. In the setting of symmetric matrix factorization, the matrices $\bm{X}^0$ and $\bm{Y}$ are generated according to the Markov chain $\bm{X}^0 \rightarrow \bm{W}^0\rightarrow \bm{Y}$, with
\begin{align}
    &\bm{X}^0\in\mathbb{R}^{N\times r}\overset{d}{=}P_X(\cdot)\\
    &\bm{W}^0\in\mathbb{R}^{N\times N}=\frac{1}{\sqrt{N}}\bm{X}^0\bm{K}\bm{X}^{0 T}\\
    &\bm{Y}\in\mathbb{R}^{N\times N}\overset{d}{=}P_{\mathrm{out}}(\cdot|\bm{W}^0),
\label{eq:matrix_facto_pb}
\end{align}
where $\bm{K}\in\mathcal{S}_{\mathbb{R}}^{r\times r}$ is a given symmetric matrix \cite{Thibault_MMSE}, and the thermodynamic limit $N \to \infty,  r=\mathcal{O}(1)$ is assumed. For the particular case of an SBM, the matrix $\bm{X}^0$ is the horizontal stack of the indicator vectors $\bm{x}_i$ for each individual $i$, with $(\bm{x}_i)_q=1$ iff $i$ belongs to community $q$, and $0$ otherwise. The prior $P_X(\cdot)$ is assumed to be row-wise separable (which corresponds to independently assigned communities in the SBM), i.e. to factorize on the rows $\{\bm{x}_i^0\}_{i=1}^{N}$ of $\bm{X}^0$,
\begin{equation}
\label{eq:prior}
P_{\bf X}(\bm{X}^0)=\prod\limits_i^NP_X(\bm{x}^0_i).
\end{equation}
The posterior $P_{\mathrm{out}}(\cdot|\bm{W})$ is assumed to factorize on all matrix components (note that, for simplicity, the same notation is kept)
\begin{equation}
P_{\mathrm{out}}(\bm{Y}|\bm{W})=\prod\limits_{i,j}P_{\rm{out}}(Y_{ij}|W_{ij}).
\label{eq:posterior}
\end{equation}
This assumption corresponds in the context of the SBM to edges being generated independently. The inference problem is to estimate the ground truth value of $\X$, $\bm{X}^0$ (up to a permutation of the columns), that was used to generate $\Y$ with the sole knowledge of $\bm{Y}$ and of the parameter $\bm{K}$. This can naturally be carried out using the Bayes rule, considering the posterior measure $P(\X|\Y)$. In the language of statistical physics, this probability defines a Hamiltonian 
\begin{equation}
\label{eq:hamiltonian}
   \mathcal{H}_{\Y}(\X)=-\ln P_{\bf X}(\X)- \ln P_{\rm out}(\Y | \W)+\ln P(\Y)
\end{equation}
see \cite{Decelle2011AsymptoticAO,Thibault_Constrained,Thibault_MMSE}. The rows of $\X$ play the role of \textit{spin variables}, while $\Y$ acts as a quenched disorder. The Gibbs measure associated with Hamiltonian \eqref{eq:hamiltonian} is 
\begin{equation}
    \mu_{\Y}(\X)=\frac{1}{Z_{\Y}}e^{-\mathcal{H}_{\Y}(\X)},
\end{equation}
the normalization factor being called the \textit{partition function}. The \textit{quenched free energy} associated is defined as 
\begin{align}
\label{eq:free_energy_matrixfacto}
    f&=-\mathbb{E}_{\X^0,\Y}\ln Z_{\Y}\nonumber\\
    &=-\mathbb{E}_{\X^0,\Y}\ln\int d\X e^{-\mathcal{H}_{\Y}(\X)}.
\end{align}

\subsection{Large deviations of the free energy under partial disclosure of the ground truth}
We are interested in the deviations of the free energy \eqref{eq:free_energy_matrixfacto} when certain rows of the ground truth $\X^0$, $\{\x_i^0\}_{i\in\SSel}$, are disclosed. We denote with $\SSel$ a subset of $\llbracket 1,N\rrbracket$ of given cardinal $|\SSel|=nN$. This is tantamount to replacing the prior \eqref{eq:prior} by
\begin{equation}
    P^{\SSel}_{\bf X}(\X)=\prod\limits_i^N\Big(\sigma_i \delta(\x_i-\x^0_i)+(1-\sigma_i)P_X(\x_i)\Big),
\label{eq:S_prior}
\end{equation}
where $\sigma_i=\mathbbm{1}_{i\in\SSel}$ is the indicator function of $i$ pertaining to the disclosed rows in $\SSel$. The goal of the present work is to evaluate, given a certain cardinal $n=|\SSel|/N$ and a given value of free energy $f$, the \textit{subset-entropy} $\Sigma(n,f)$, i.e. the logarithmic number of ways to select a subset $\SSel$ of intensive cardinal $n$ so that the free energy is equal to $f$. While obtaining directly this quantity is not straightforward, a viable method for its evaluation is by means of a Legendre transform \cite{obuchi2018statistical, Cui2020LargeDF}. First, we evaluate the grand canonical free entropy
\begin{align}
\label{eq:LD_principle}
    &\Phi(\beta,\phi)\nonumber\\
    &=\mathbb{E}_{\X^0,\Y}\ln\sum\limits_{\{\sigma_i\}_{i=1}^N}\Bigg[
    \int d\X P^{\SSel}_X(\X)\frac{P_{\mathrm{out}}(\Y|\W)}{P_{\mathrm{out}}(\Y|\bm{0})}
    \Bigg]^\beta e^{\phi \sum\limits_i^N\sigma_i}.
\end{align}
In \eqref{eq:LD_principle} $\beta$ plays the role of a (possibly negative) inverse temperature, and $\phi$ is a chemical potential controlling the cardinal of the subset. As mentioned before, the components of $\Y$ act as quenched disorder, while selection variables $\sigma$ act as annealed disorder, whose fluctuation timescale differs from the annealed spins $\x_i$ by a factor controlled by $\beta$. In high dimensions, the free-entropy $\Phi$ is expected to concentrate:
\begin{equation}
    \Phi(\beta,\phi)=\underset{n,f}{\rm{extr}}~\Sigma(n,f)-\beta f +\phi n,
\end{equation}
where $f,n$ are the dominant values of free-energy and cardinality at the given $\beta$ and $\phi$. 
Inverting the Legendre transform yields the sought subset-entropy function, as 
\begin{equation}
    \Sigma(n,f)=\underset{\beta,\phi}{\rm{extr}}~\Phi(\beta,\phi)+\beta f -\phi n.
\end{equation}

\section{Replica computation} \label{sec:replica}

\subsection{Replica trick}
The free entropy \eqref{eq:LD_principle} can be computed using the replica method \cite{Replica,Replica2,MezardVirasoro}. Calling $\Xi$ the argument of the logarithm in \eqref{eq:LD_principle}, a simple mathematical identity allows one to rewrite the expression as:
\begin{equation}
\label{eq:replica_trick}
    \mathbb{E}_{\X^0,\Y}\ln \Xi=\underset{s\rightarrow0}{\rm{lim}}\frac{1}{s}\ln\mathbb{E}_{\X^0,\Y}\Xi^s.
\end{equation}
Assuming an integer value of $\beta$, the free entropy \eqref{eq:LD_principle} can then be fully rewritten in terms of replicas of the original system:
\begin{widetext}
\begin{align}
\label{eq:repl_entro}
\Phi(\beta,\phi)
&=\underset{s\rightarrow 0}{\mathrm{lim}}
\frac{1}{s}\int\prod\limits_{i=1}^{N}d\bm{x}^{0}_{i}P_{X}(\bm{x}^{0}_{i})
\prod\limits_{i\leq j}dY_{ij} P_{\mathrm{out}}(Y_{ij}|0)e^{g(Y_{ij}|W^{0}_{ij})-g(Y_{ij}|0)}
\nonumber\\
&\sum\limits_{\{\sigma_{i}^{a}\}}
\int\prod\limits_{a=1}^{s}\prod\limits_{\alpha=1}^{\beta}\prod\limits_{i=1}^{N}d\bm{x}_{i}^{a\alpha}
((1-\sigma_{i}^{a})P_{X}(\bm{x}_{i}^{a\alpha})+\sigma_{i}^{a}\delta(\bm{x}_{i}^{a\alpha}-\bm{x}^{0}_{i}))
\prod\limits_{i\leq j}e^{\sum\limits_{a\alpha}g(Y_{i j}|W_{i j}^{a\alpha})-g(Y_{i j}|0)}
e^{\phi\sum\limits_{a,i}\sigma_{i}^{a}}.
\end{align}
\end{widetext}
Following \cite{Thibault_Constrained,Thibault_MMSE}, we have adopted the notation 
\begin{equation}
    g(\Y|\W)\overset{\rm{def}}{=}\ln P_{\rm{out}}(\Y|\W),
\end{equation}
with $g$ thereby corresponding to the spin-spin interaction part of the Hamiltonian \eqref{eq:hamiltonian}.
Note the presence of two levels of replications: the first replica index $a$ runs over $\llbracket 1, s\rrbracket$ and corresponds to the replication introduced by the trick \eqref{eq:replica_trick}. The second replication index $\alpha$ is on the other hand induced by the $\beta$ power in \eqref{eq:LD_principle}. Note that, at the end of the computation we will restore the initial continuous temperature via an analytical continuation to any value $\beta\in\mathbb{R}$.
For the sake of compactness, we index the ground truth spins $\{\x_i^0\}_{i=1}^N$ with the replica index $a=0$ (and set accordingly $\sigma_i^0=0$ for all $i$ for coherence). We also denote the 
effective prior $((1-\sigma_{i}^{a})P_{X}(\bm{x}_{i}^{a\alpha})+\sigma_{i}^{a}\delta(\bm{x}_{i}^{a\alpha}-\bm{x}^{0}_{i}))$ as $P^{\sigma^a_i}_X(\x_i^{a\alpha})$, with $\SSel_a$ the $a^{\rm{th}}$ replica of subset $\SSel$. \\

\subsection{Channel universality}
  Expanding the interaction term up to $\mathcal{O}\left(\frac{1}{N}\right)$ we get:
\begin{align}
    &e^{\sum\limits_{a\alpha}g(Y_{ij}|W_{ij}^{a\alpha})-g(Y_{ij}|0)}\nonumber\\
    &=1+\partial_{W}g(Y_{ij}|0)
    \sum\limits_{a\alpha}W_{ij}^{a\alpha}+\frac{1}{2}\partial_{W}^{2}g(Y_{ij}|0)
    \sum\limits_{a\alpha}(W_{ij}^{a\alpha})^{2}\nonumber\\
    &
    +\frac{1}{2}(\partial_{W}g(Y_{ij}|0))^{2}
    \sum\limits_{a\alpha,c\gamma}W_{ij}^{a\alpha}W_{ij}^{c\gamma}
    +o\Bigg(\frac{1}{N}\Bigg).
\end{align}
It can be further shown \cite{Thibault_Constrained,Thibault_MMSE} that
\begin{align}
&\int P_{\mathrm{out}}(y)dy\partial_{W}g(Y_{ij}|0)=0\\
&\int P_{\mathrm{out}}(y)dy(\partial^{2}_{W}g(Y_{ij}|0)
+\partial_{W}g(Y_{ij}|0)^2)=0.
\end{align}
Moreover, defining the effective noise also known as \textit{Fisher score} $\Delta$ as 
\begin{align}
    \frac{1}{\Delta}=\mathbb{E}_{P(.|0)}((\partial_{W}g(Y_{ij}|0))^{2}),
\end{align}
the replica free entropy \eqref{eq:repl_entro} can be cast into the form
\begin{align}
\label{eq:repl_entro2}
\Phi(\beta,\phi)=&
\sum\limits_{\{S_{i}^{a}\}}
\int\prod\limits_{a\alpha}\prod\limits_{i=1}^{N}d\bm{x}_{i}^{a\alpha}
P^{\sigma^a_i}_{X}(\bm{x}_{i}^{a\alpha})\nonumber\\
&e^{\frac{1}{2\Delta}\sum\limits_{i,j}\sum\limits_{a\alpha<c\gamma}W_{ij}^{a\alpha}W_{ij}^{c\gamma}}
e^{\phi\sum\limits_{a,i}\sigma_{i}^{a}}.
\end{align}
Note that all dependence with respect to the output measure $P_{\rm{out}}$ has been subsumed into the effective noise $\Delta$. In previous work, this property has been referred to as \textit{channel universality} \cite{Thibault_Constrained,Thibault_MMSE,review}, and allows connections between apparently disparate inference problems.

\subsection{Casting as a saddle-point problem}
It is possible to rewrite the computation of $\Phi(\beta,\phi)$ as a saddle-point problem. To that end, define the overlap tensor
\begin{align}
\label{eq:def_Q}
    \T_{a\alpha,c\gamma}=\frac{1}{N}\sum\limits_{i=1}^{N}\bm{x}_{i}^{a\alpha}
    (\bm{x}_{i}^{c\gamma})^{T} \in (\mathbb{R}^{r\times r})^{(s\times\beta)\times (s\times\beta)}.
\end{align}
$\T$ is a $(s\times\beta)\times (s\times\beta)$ block matrix with blocks in $\mathbb{R}^{r\times r}$. Arguably, the overlap matrix is mathematically ill defined outside of the replica trick setup, as $s\rightarrow 0$ and $\beta$ is a (potentially negative) real number.
A notable property of $\T$ following from its definition \eqref{eq:def_Q} is its symmetry up to transposition with respect to its first two indices,
\begin{align}
    \T_{a\alpha,c\gamma}=(\T_{c\gamma,a\alpha})^{T}.
\end{align}
This property allows one to compactly rewrite the interaction Hamiltonian as 
\begin{align}
    \sum\limits_{i,j}\sum\limits_{a\alpha<c\gamma}W_{ij}^{a\alpha}W_{ij}^{c\gamma}=N\sum\limits_{a\alpha<c\gamma}
    \Tr(\T_{a\alpha,c\gamma}\bm{K}\T_{c\gamma,a\alpha}\bm{K}).   
\end{align}
The final step in order to rewrite \eqref{eq:repl_entro2} as an extremization problem is to introduce the Fourier representation of a Dirac's delta function, enforcing definition \eqref{eq:def_Q}:
\begin{align}
    1=&\int\prod\limits_{a\alpha\leq c\gamma}d\T_{a\alpha,c\gamma}d\bm{\hat{\T}}_{a\alpha,c\gamma}\nonumber\\
    &
    e^{-N\sum\limits_{a\alpha\leq c\gamma}\mathrm{tr}\left(\hat{\T}_{a\alpha,c\gamma}
    (\T_{a\alpha,c\gamma}-\frac{1}{N}\sum\limits_{i=1}^{N}\bm{x}_{i}^{a\alpha}
    \bm{x}_{i}^{c\gamma T})^T\right)}.
\end{align}
We have absorbed the imaginary factor $i$ into the definition of the Fourier conjugate $\hat{\T}$ and omitted some subdominant terms induced by the normalization of the Dirac Fourier transform in the exponential.
Then \eqref{eq:repl_entro2} can be compactly rewritten as

\begin{align}
\label{eq:SP_form_phi}
    \Phi(\beta,\phi)
    =
    \int\prod\limits_{a\alpha\leq c\gamma}d\T_{a\alpha,c\gamma}d\bm{\hat{\T}}_{a\alpha,c\gamma}\nonumber
    e^{N\ln I(\bm{\hat{\T}})+NT(\T,\hat{Q)}}
\end{align}
with
\begin{align}
    &I(\bm{\hat{\T}})\nonumber\\
    &=\sum\limits_{\sigma^{a}}
        \int\prod\limits_{a\alpha}d\bm{x}^{a\alpha}
    P^{\sigma^a}_{X}(\bm{x}^{a\alpha})e^{\sum\limits_{a\alpha\leq c\gamma}\mathrm{tr}\left(\hat{\T}_{a\alpha,c\gamma}
    \bm{x}^{c\gamma}
    \bm{x}^{a\alpha T}\right)}
    e^{\phi\sum\limits_{a}\sigma^{a}} 
\end{align}
and
\begin{align}
T(\T,\hat{\T})&=\frac{1}{2\Delta}\sum\limits_{a\alpha<c\gamma}
    \Tr(\T_{a\alpha,c\gamma}\bm{K}\T_{c\gamma,a\alpha}\bm{K})\nonumber\\
    &-\sum\limits_{a\alpha\leq c\gamma}\mathrm{tr}(\hat{\T}_{a\alpha,c\gamma}
    \T_{a\alpha,c\gamma}^T).
\end{align}
Equation \eqref{eq:SP_form_phi} suggests to employ a saddle-point type approximation to compute the free entropy $\Phi(\beta,\phi)$ as 

\begin{align}
\label{eq:extr_pb}
    \Phi(\beta,\phi)=\underset{\T,\bm{\hat{\T}}}{\mathrm{extr}} \left[T(\T,\hat{\T})
    +\ln I(\bm{\hat{\T}})\right],
\end{align}
with the extremization carried out over the matrix $\T$ and its Dirac conjugate $\hat{\T}$.

\subsection{RS ansatz}
As written in \eqref{eq:extr_pb}, the extremization must be carried out in $((\mathbb{R}^{r\times r})^{(s\times\beta)\times (s\times\beta)})^2$ and is nor mathematically well-defined nor tractable. Progress can be made  \cite{MezardVirasoro,Replica,Replica2} by making an assumption on the form of the overlap matrices extremizing \eqref{eq:SP_form_phi}. We here assume the simplest, and most common ansatz, the \textit{Replica Symmetric} (RS) ansatz. The RS ansatz in our case is readily amenable to physical interpretation. Assuming replica symmetry in the large deviations setting amounts to assuming that any two variables seeing the 
same disorder $\{\sigma_i^a\}_{i=1}^N$ (i.e. same $a$ index) have the same $r\times r$ overlap,
while two seeing different disorders (viz. different $a$ index)
possess a distinct and unique overlap:
\begin{align}
\label{eq:ansatz_begin}
&\T^{0,0}=\bm{r}^{0},&& \bm{\hat{\T}}^{0,0}=\bm{\hat{r}}^{0}\\
&\T^{a\alpha,a\alpha}=\bm{r},&& \bm{\hat{\T}}^{a\alpha,a\alpha}=\bm{\hat{r}}\\
&\T^{0,a\alpha}=\bm{m},&& \bm{\hat{\T}}^{0,a\alpha}=\bm{\hat{m}}\\
&\T^{a\alpha,a\gamma}=\bm{Q},&& \bm{\hat{\T}}^{a\alpha,a\gamma}=\bm{\hat{Q}}\\
\label{eq:ansatz_end}
&\T^{a\alpha,c\gamma}=\bm{q},&& \bm{\hat{\T}}^{a\alpha,c\gamma}=\bm{\hat{q}}
\end{align}
This ansatz can be simplified by setting the gradient with respect to components of $\T$ to $0$ in \eqref{eq:SP_form_phi} $\partial_{\T_{a\alpha,c\gamma}^{\mu\nu}}\Phi{=}0$. Since all dependence on $\T$ in $\Phi(\beta,\phi)$ is contained in $T(\T,\hat{\T})$, this saddle point condition implies
\begin{equation}
    \partial_{\T_{a\alpha,c\gamma}^{\mu\nu}}T(\T,\hat{\T})=\mathbbm{1}_{a\alpha\ne c\gamma}\frac{1}{\Delta}(\K\T_{a\alpha,c\gamma}\K)^{\mu\nu}
    -\hat{\T}_{a\alpha,c\gamma}^{\mu\nu}{=}0.
\end{equation}
So the components of the conjugate tensor $\hat{\T}$ are actually related to those of the overlap tensor $\T$ by
\begin{align}
\label{eq:hat_relations}
    &\hat{\bm{r}}^{0}=\hat{\bm{r}}=0,\\
    &\hat{\bm{m}}=\frac{1}{\Delta}\bm{K}\bm{m}\bm{K},\\
    &\hat{\bm{Q}}=\frac{1}{\Delta}\bm{K}\bm{Q}\bm{K},\\
    &\hat{\bm{q}}=\frac{1}{\Delta}\bm{K}\bm{q}\bm{K}.
\end{align}
Note that, in Bayes-optimal inference, replica symmetry is usually guaranteed to hold by the Nishimori identities \cite{Nishimori}. However, these identities only apply to the equilibrium solutions of the system, obtained at $\beta=0$ in our framework. Therefore this simplifying RS ansatz needs to be validated through a stability analysis (presented in section \ref{sec:stability}).

\subsection{Replica free entropy}
\label{subsection:computationRS}
The term $T(\T,\hat{\T})$ can be evaluated explicitly using the RS ansatz \eqref{eq:ansatz_begin}-\eqref{eq:ansatz_end} and using the relations \eqref{eq:hat_relations}
\begin{align}
    T(\T,\hat{\T})=&-s\frac{\beta }{2\Delta} \Tr(\bm{m}\bm{K}\bm{m}^{T}\bm{K})
    -s\frac{\beta(\beta-1)}{4\Delta}\Tr(\bm{Q}\bm{K}\bm{Q}^{T}\bm{K})\nonumber\\
    &
    -\frac{\beta^{2}}{2\Delta}\frac{s(s-1)}{2}\Tr(\bm{q}\bm{K}\bm{q}^{T}\bm{K})
\end{align}

We can expand the exponent in $I(\bm{\hat{\T}})$ as follows:
\begin{align}
    &e^{\sum\limits_{a\alpha\leq c\gamma}\mathrm{tr}\left(\hat{\T}_{a\alpha,c\gamma}
    \bm{x}^{c\gamma}
    \bm{x}^{a\alpha T}\right)}\nonumber\\
    &=\mathbb{E}^{\mathcal{N}(0,1)}_{\bm{\xi}}\prod\limits_a \mathbb{E}^{\mathcal{N}(0,1)}_{\bm{\lambda}_a}e^{(\bm{x}^{0})^{T}\bm{\hat{m}}\sum\limits_{a\alpha\ne0}\bm{x^{a\alpha}}
    -\frac{1}{2}\sum\limits_{a\alpha\ne0}(\bm{x^{a\alpha}})^{T}
    \bm{\hat{Q}}\bm{x^{a\alpha}}}\nonumber\\
    &\times e^{+\sum\limits_{a\ne0}\lambda_{a}(\hat{\bm{Q}}-\hat{\bm{q}})^{\frac{1}{2}}
    \sum\limits_{\alpha}\bm{x^{a\alpha}}+\bm{\xi}^{T}\bm{\hat{q}}^{\frac{1}{2}}\sum\limits_{a\alpha\ne0}x^{a\alpha}}.
\end{align}
We introduced two Hubbard-Stratonovitch
fields $\bm{\xi}$ and $\{\bm{\lambda_{a}}\}_{a=1}^{s}$ to linearise the exponent.
\begin{align}
    &I(\bm{\hat{\T}})
    =\sum\limits_{\sigma^{a}}\mathbb{E}^{\mathcal{N}(0,1)}_{\bm{\xi}}\prod\limits_a \mathbb{E}^{\mathcal{N}(0,1)}_{\bm{\lambda}_a}
    \prod\limits_{a\alpha}d\bm{x}^{a\alpha}
    P_{X}^{\sigma_a}(\bm{x}^{a\alpha})\nonumber\\
    & \times e^{(\bm{x}^{0})^{T}\bm{\hat{m}}\sum\limits_{a\alpha\ne0}\bm{x^{a\alpha}}
        -\frac{1}{2}\sum\limits_{a\alpha\ne0}(\bm{x^{a\alpha}})^{T}
        \bm{\hat{Q}}\bm{x^{a\alpha}}}\nonumber\\
    &
       \times e^{+\sum\limits_{a\ne0}\lambda_{a}(\hat{\bm{Q}}-\hat{\bm{q}})^{\frac{1}{2}}
        \sum\limits_{\alpha}\bm{x^{a\alpha}}
        +\bm{\xi}^{T}\bm{\hat{q}}^{\frac{1}{2}}\sum\limits_{a\alpha\ne0}x^{a\alpha}}\nonumber\\
    &=\mathbb{E}^{\mathcal{N}(0,1)}_{\bm{\xi}}\mathbb{E}^{P_X}_{\bm{\x^0}}\nonumber\\
        &\Bigg[\sum\limits_{\sigma=0,1}e^{\phi \sigma}\mathbb{E}^{\mathcal{N}(0,1)}_{\bm{\lambda}} 
        \Bigg\{
            \int d\bm{x}((1-\sigma)P_{X}(\bm{x})+\sigma\delta(\bm{x}-\bm{x}^{0}))\nonumber\\
    &
            e^{-\frac{1}{2}\bm{x}^{T}\hat{\bm{Q}}\bm{x}
            +((\bm{x}^{0})^{T}\bm{\hat{m}}+\bm{\lambda}^{T}
            (\bm{\hat{Q}}-\bm{\hat{q}})^\frac{1}{2}+\bm{\xi}^{T}
            \bm{\hat{q}}^\frac{1}{2})\bm{x}}
        \Bigg\}^{\beta}
        \Bigg]^{s}.
\end{align}

This allows to explicitly write the free entropy \eqref{eq:repl_entro} as an extremization problem on $\mathbb{R}^{r\times r}$. Using the shorthand
\begin{align}
    h(\bm{x})\overset{\rm{def}}{=}-\frac{1}{2}\bm{x}^{T}\hat{\bm{Q}}\bm{x}
    +((\bm{x}^{0})^{T}\bm{\hat{m}}+\bm{\lambda}^{T}
    (\bm{\hat{Q}}-\bm{\hat{q}})^\frac{1}{2}+\bm{\xi}^{T}
    \bm{\hat{q}}^\frac{1}{2})\bm{x},
\end{align}
the free entropy is given by
\begin{align}
 \label{eq:final_entro}
    \Phi(\beta,\phi)&=\underset{\bm{m},\bm{Q},\bm{q}}{\mathrm{extr}}~~-\frac{\beta}{2\Delta} \Tr(\bm{m}\bm{K}\bm{m}^{T}\bm{K})
    \nonumber\\
    &-\frac{\beta(\beta-1)}{4\Delta}\Tr(\bm{Q}\bm{K}\bm{Q}^{T}\bm{K})
    +\frac{\beta^{2}}{4\Delta}\Tr(\bm{q}\bm{K}\bm{q}^{T}\bm{K})
    \nonumber\\
    &
    +\mathbb{E}^{\mathcal{N}(0,1)}_{\bm{\xi}}\mathbb{E}^{P_X}_{\bm{\x^0}}\ln
    \Bigg[\mathbb{E}^{\mathcal{N}(0,1)}_{\bm{\lambda}}e^{\phi}e^{\beta h(\bm{x}^{0})}\nonumber\\
    &+
    \mathbb{E}^{\mathcal{N}(0,1)}_{\bm{\lambda}}\Bigg\{
    \int d\bm{x}\frac{P_{X}(\bm{x})}{P_{X}(\bm{x}^{0})}e^{h(\bm{x})}
    \Bigg\}^{\beta}
    \Bigg].
\end{align}

\subsubsection{Saddle point equations}
The extremization in \eqref{eq:final_entro} can be carried out by explicitly writing the zero-gradient conditions on each of the order parameters $\bm{m},\Q,\bm{q}$ and iterating them until convergence. The zero-gradient condition are in the general case hard to compute, since they involve matrix derivatives. Though we believe that progress can be made in some particular cases where the matrix structure of the order parameters is known (for example if the prior is invariant over the permutation group $\mathfrak{S}(\llbracket 1,r\rrbracket)$ ), we restrict ourselves in this manuscript to the scalar $r=1$ case and leave the generic matrix gradient derivation to future work. For $r=1$, the saddle-point equations read:

\begin{align}
    &h(x)-\frac{Qx^2}{2\Delta}
    +\frac{x}{\Delta^{\frac{1}{2}}}\left(\frac{
x^{0}m}{\Delta^{\frac{1}{2}}}+\lambda
    (Q-q)^\frac{1}{2}+\xi
    q^\frac{1}{2}\right)\label{eq:SP1}\\
    &Z_x=\int dxP_{X}(x)e^{h(x)}\\
    &f_x=\frac{1}{Z_x}\int dx x P_{X}(x)e^{h(x)}\\
    &m=\mathbb{E}^{\mathcal{N}(0,1)}_{\xi}\mathbb{E}^{P_X}_{x^0}
    \frac{ \mathbb{E}^{\mathcal{N}(0,1)}_{\lambda}\Big[e^{\phi}(x^{0})^{2}e^{\beta h(x^{0})}
    +Z_x^{\beta}
    f_xx^{0}        
    \Big]}
    {\mathbb{E}^{\mathcal{N}(0,1)}_{\lambda}\Big[e^{\phi}e^{\beta h(x^{0})}+
    Z_x^{\beta}\Big]}\\
    &Q=\mathbb{E}^{\mathcal{N}(0,1)}_{\xi}\mathbb{E}^{P_X}_{x^0}
    \frac{\mathbb{E}^{\mathcal{N}(0,1)}_{\lambda}\Big[e^{\phi}(x^{0})^2e^{\beta h(x^{0})}
    +Z_x^{\beta}
    f_x^2  
    \Big]}
    {\mathbb{E}^{\mathcal{N}(0,1)}_{\lambda}\Big[e^{\phi}e^{\beta h(x^{0})}+
    Z_x^{\beta}\Big]}\\
    &q=\mathbb{E}^{\mathcal{N}(0,1)}_{\xi}\mathbb{E}^{P_X}_{x^0}
    \Bigg(\frac{\mathbb{E}^{\mathcal{N}(0,1)}_{\lambda}\Big[e^{\phi}x^{0}e^{\beta h(x^{0})}
    +Z_x^{\beta}
    f_x       
    \Big]}
    {\mathbb{E}^{\mathcal{N}(0,1)}_{\lambda}\Big[e^{\phi}e^{\beta h(x^{0})}+
    Z_x^{\beta}\Big]}\Bigg)^2
    \label{eq:SP2}
\end{align}

\paragraph{Saddle point equations for the SBM with two balanced communities}
We give the specialization of the saddle-point equations \eqref{eq:SP1}-\eqref{eq:SP2} for the SBM with two balanced communities defined by \eqref{eq:SBM_prior} and \eqref{eq:SBM_posterior}.
The effective noise $\Delta$ in this case is \cite{Thibault_Constrained,Thibault_MMSE}
\begin{align}
    \Delta=\frac{p(1-p)}{\mu^2}.
\end{align}
Since the SBM with two balanced communities is a particular case of $r=1$ factorization, all order parameters are scalars. Define the quantities
\begin{align}
    & Z_x= \rho e^{h_+}+(1-\rho) e^{h_-}\label{eq:shorthand_SBM_1}\\
    &f_x=\frac{1}{Z_x}( \rho x_+ e^{h_+}+(1-\rho) x_- e^{h_-})\\
    &h_+=-\frac{1}{2\Delta}Qx_+^2+\Big(\frac{mx^0}{\Delta}+\frac{\lambda\sqrt{Q-q}}{\sqrt{\Delta}}+\frac{\xi \sqrt{q}}{\sqrt{\Delta}}\Big)x_+\\
    &h_-=-\frac{1}{2\Delta}Qx_-^2+\Big(\frac{mx^0}{\Delta}+\frac{\lambda\sqrt{Q-q}}{\sqrt{\Delta}}+\frac{\xi \sqrt{q}}{\sqrt{\Delta}}\Big)x_-\\
    &\beta h_0=-\frac{\beta}{2\Delta}Q(x^0)^2+\beta\Big(\frac{mx^0}{\Delta}+\frac{\xi \sqrt{q}}{\sqrt{\Delta}}\Big)x^0+\frac{Q-q}{2\Delta}\beta^2(x^0)^2\label{eq:shorthand_SBM_2}
\end{align}
For readability, we dropped in the notations of $Z_x, f_x, h_\pm, h_0$ the dependence on the order parameters $q, Q, m$, $x^0$ and the Hubbard-Stratonovitch fields $\xi,\lambda$.
Using these shorthands, the updates read 

\begin{align}
    &m=\mathbb{E}^{\mathcal{N}(0,1)}_{\xi}\mathbb{E}^{P_X}_{x^0}\frac{e^{\phi+\beta h_0}(x^0)^2+\mathbb{E}^{\mathcal{N}(0,1)}_{\lambda} Z_x^\beta f_xx^0}{e^{\phi+\beta h_0}+\mathbb{E}^{\mathcal{N}(0,1)}_{\lambda} Z_x^\beta},\\
    &Q=\mathbb{E}^{\mathcal{N}(0,1)}_{\xi}\mathbb{E}^{P_X}_{x^0}\frac{e^{\phi+\beta h_0}(x^0)^2+\mathbb{E}^{\mathcal{N}(0,1)}_{\lambda} Z_x^\beta f_x^2}{e^{\phi+\beta h_0}+\mathbb{E}^{\mathcal{N}(0,1)}_{\lambda} Z_x^\beta},\\
    &q=\mathbb{E}^{\mathcal{N}(0,1)}_{\xi}\mathbb{E}^{P_X}_{x^0}\Bigg(\frac{e^{\phi+\beta h_0}x^0+\mathbb{E}^{\mathcal{N}(0,1)}_{\lambda} Z_x^\beta f_x}{e^{\phi+\beta h_0}+\mathbb{E}^{\mathcal{N}(0,1)}_{\lambda} Z_x^\beta }\Bigg)^2.
\end{align}

\begin{figure}
    \centering
    \includegraphics[scale=.6]{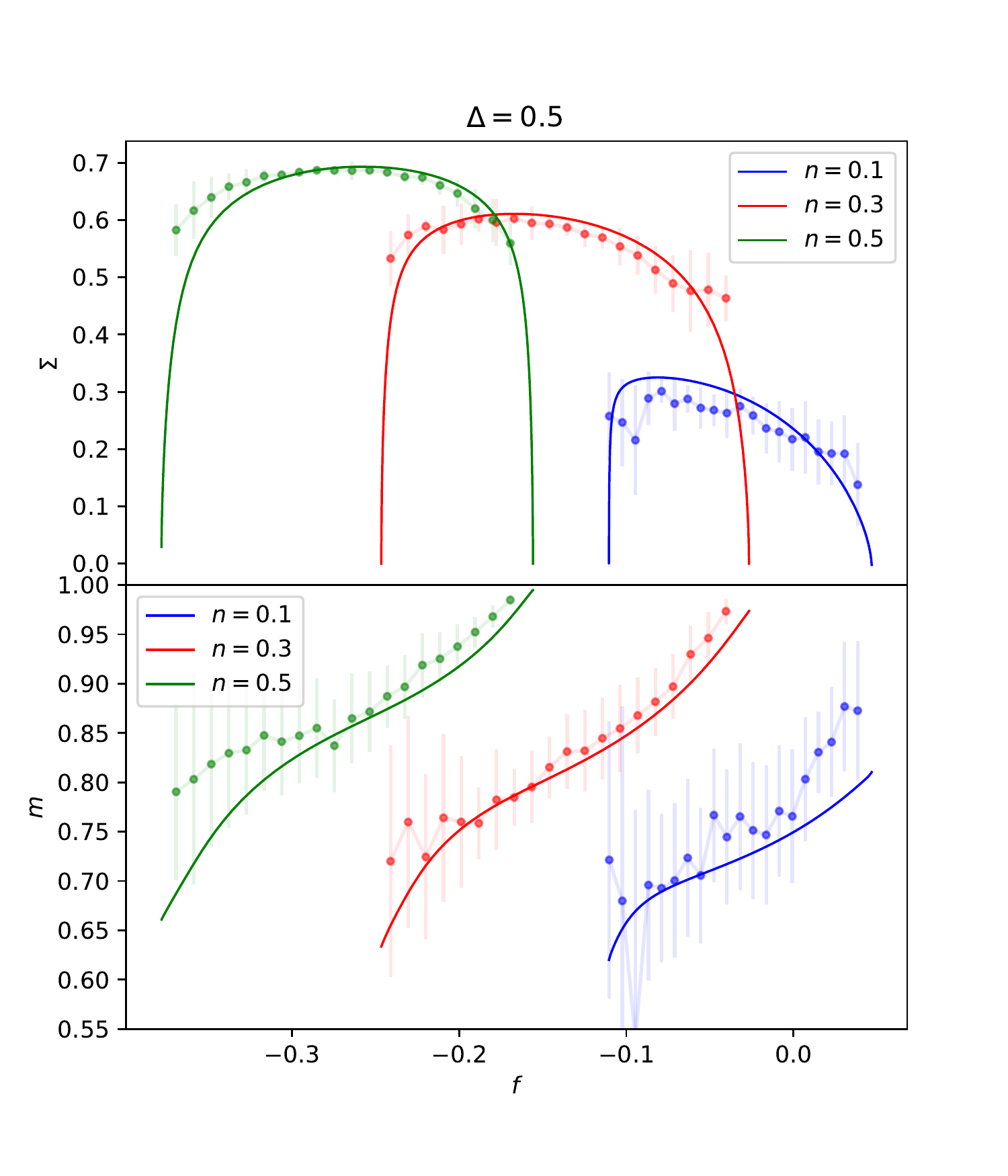}\\
    \vspace{-1cm}
    \includegraphics[scale=.6]{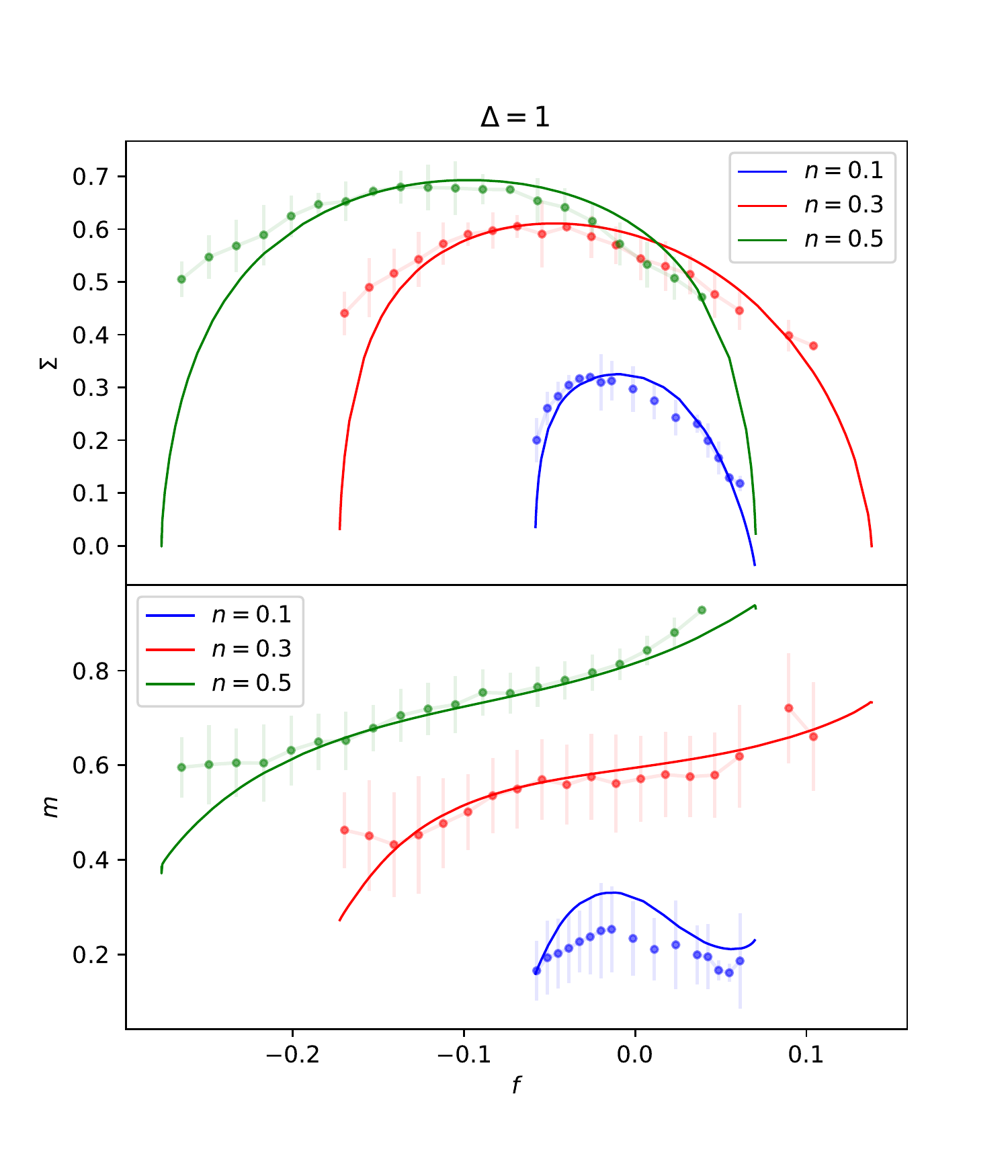}
    \caption{Subset-entropy $\Sigma(n,f)$ (upper) and typical magnetization $m(n,f)$ (lower) for small effective noise $\Delta=0.5$ (top plot) and critical effective noise $\Delta=1$ (bottom plot), and disclosed fractions $n\in\{0.1,~0.3,~0.5\}$. Solid lines correspond to asymptotic predictions in the large $N$ limit as given by the large deviations. Dashed lines correspond to simulations averaged over $20$ instances of SBM with $N=50$ nodes, by repeatedly drawing for each instance subsets of intensive size $n$ at random $10^7$ times and evaluating $f,m$ for each draw using the TAP \cite{TAP,review} algorithm. }
    \label{fig:Sigma_curves}
\end{figure}

\section{Results}\label{sec:results}

In this section we display the results of the presented large deviation analysis. In section \ref{sec:replica}, starting from a  high-dimensional problem, we derived an analytical expression for the free-entropy that only depends on a small $\mathcal{O}(1)$ set of overlap parameters. The saddle-point values of these parameters, obtained by extremizing the free-entropy, can now be used to predict relevant metrics of the inference performance in large SBM instances. 

In particular, through a Legendre transform, we can look at the free-energy/subset-entropy curves, spanning among all possible choices for the subset of revealed nodes at fixed cardinality $n$. The subset-entropy is linked to the rareness of the correlated subset selections, and can be interpreted as a measure of the probability of randomly sampling a subset with given $n,f$ \cite{Cui2020LargeDF}. The maxima of the curves represent the typical choice of revealed nodes, and the associated value of the log-posterior in the semi-supervised inference problem. In high dimensions, all fluctuations around the typical free energy are exponentially suppressed and thus extremely unlikely in large instances. In Fig.~\ref{fig:Sigma_curves}, upper panels, we show the analytic curves, compared with the numerical results in dimension $N=50$ (the small size allows rare events to be observed in reasonable simulation time, with $10^7$ samples). Note that negative values of the subset-entropy are completely unphysical, and signal a problem in the assumptions made in the replica computation. We look further into this problem in section \ref{sec:stability}. 

Since the free-energy is not immediately interpretable as a performance measure in the SBM setting, we need a different way of characterizing the difficulty of the semi-supervised learning problem associated to a given set of labelled nodes. A natural measure of the helpfulness of disclosing a set $\mathcal{S}$ is the mean reconstruction error, i.e. the overlap of the inferred rows of $\bm{X}$ with those of the ground truth $\bm{X}^0$, with respect to the Bayes posterior measure $P(\bm{X}|\bm{Y})$,
\begin{equation}
    \mathrm{tr}\, \bm{m}=\frac{1}{N}\sum\limits_{i=1}^N \mathbb{E}_{\bm{x}}^{P(\bm{X}|\bm{Y})}\left[\bm{x}_i^T\bm{x}_i^0\right]. \label{eq:reconstruction}
\end{equation}
Note that for rank one problems, $\mathrm{tr}\, \bm{m}$ just reduces to $m$. Smart choices for $\mathcal{S}$ lead to easier inference problems and therefore, on average, to better inference of the planted community assignment (signalled by a larger mean overlap). 

In the lower panels in Fig.~\ref{fig:Sigma_curves}, we show the reconstruction overlap obtained in correspondence of different revealed subsets, and the corresponding numerical simulations at $N=50$. A first observation is that very rare node selections can lead to much higher information gains, which justifies the implementation of algorithmic strategies for their identification.
Surprisingly, we also find that the relationship between the free-energy associated to the posterior measure of the inference problem and $m$ is not always monotonous, depending on the value of the SNR. In particular, at high SNR (small noise $\Delta=0.5$) the overlap with the signal increases with the free energy for all fraction of disclosed nodes $n$. However, when the SNR approaches the critical value ($\Delta=1$), the subset selections with the highest possible free energy are no longer associated with a better reconstruction. Note that this non-monotonous dependence is not an artifact of the RS replica analysis, as it is observed even for ranges of $n$ where the RS fixed point \eqref{eq:final_entro} is stable, see section \ref{sec:stability}, and in the finite size experiments presented in section \ref{sec:consequences}. This finding highlights the non-trivial role played by the noise in the problem of inferring the correct membership of the nodes (cfr. with the noiseless setting in \cite{Cui2020LargeDF}).

\section{Stability of the RS assumption} \label{sec:stability}

The RS ansatz \eqref{eq:ansatz_begin}-\eqref{eq:ansatz_end} is expected to hold when the configurations dominating the large deviation measure \eqref{eq:LD_principle} do not deviate too much from the typical ones. For other regions in the $\beta,\phi$ plane however, the matrices extremizing \eqref{eq:SP_form_phi} may develop more involved structure, and further steps of Replica Symmetry Breaking \cite{RSB,MezardVirasoro} might be needed. In this section, the stability of the assumption under an infinitesimal 1RSB perturbation \eqref{eq:ansatz_begin}-\eqref{eq:ansatz_end} is investigated.

For any fixed disorder $\{\sigma_i\}_{i=1}^N$, the inference problem associated to the $\beta$ replicas evolving in this disorder is Bayes optimal \cite{review} and therefore RS is expected to hold at that level of replication, i.e. $\alpha$ indices are RS. It may, however, be the case that clusters form in the $\sigma$ space, yielding a 1RSB-type landscape for the first replication level (index $a$).  The $a$ index must then be differentiated into a double $(a'\in\llbracket1,\frac{s}{\tau} \rrbracket,a\in\llbracket 1,\tau\rrbracket)$ index. The parameter $\tau$ is called in this context the \textit{Parisi parameter}. In the following, $a'$ labels the 1RSB cluster of the disorder configuration $\SSel^{a',a}$, while the $a$ index differentiates disorder configurations within a same 1RSB cluster.

\subsection{1RSB perturbation to the RS ansatz}
We consider a perturbation to the ansatz \eqref{eq:ansatz_begin}-\eqref{eq:ansatz_end} of a 1RSB form:
\begin{align}
\label{eq:perturb_begin}
&\T^{0,0,0}=\bm{r}^{0},&& \bm{\hat{\T}}^{0,0,0}=\bm{\hat{r}}^{0}\\
&\T^{a'a\alpha,a'a\alpha}=\bm{r},&& \bm{\hat{\T}}^{a'a\alpha,a'a\alpha}=\bm{\hat{r}}\\
&\T^{0,a'a\alpha}=\bm{m},&& \bm{\hat{\T}}^{0,a'a\alpha}=\bm{\hat{m}}\\
&\T^{a'a\alpha,a'a\gamma}=\Q,&& \bm{\hat{\T}}^{a'a\alpha,a'a\gamma}=\bm{\hat{Q}}\\
&\T^{a'a\alpha,a'c\gamma}=\bm{q}+\bm{\epsilon},&& \bm{\hat{\T}}^{a'a\alpha,a'a\gamma}=\bm{\hat{q}}+\hat{\eps}\\
\label{eq:perturb_end}
&\T^{a'a\alpha,c'c\gamma}=\bm{q},&& \bm{\hat{\T}}^{a\alpha,c\gamma}=\bm{\hat{q}}
\end{align}
$\eps$ is a small perturbation. All components of the matrix  $\eps$ are expected to be positive and assumed $o(1)$, since spins pertaining to the same RSB cluster should have greater overlap than spins coming from distinct clusters. Note that relations \eqref{eq:hat_relations} still hold, with the addition of a similar equation for $\eps$
\begin{equation}
    \hat{\eps}=\frac{1}{\Delta}\bm{K}\eps\bm{K}.
\end{equation}

The RS assumption of section \ref{sec:replica} is not valid if a perturbation of the form \eqref{eq:perturb_begin}-\eqref{eq:perturb_end} does cause the fixed-point defined by \eqref{eq:SP1}-\eqref{eq:SP2} to become unstable.

\subsection{1RSB free entropy}
The computation follows the same lines of the one presented above in \ref{subsection:computationRS}. At the end, one gets for the trace term $T(\T,\hat{\T})$ 
\begin{align}
    T(\T,&\hat{\T})\nonumber\\
    &=-s\frac{\beta}{2\Delta}\Tr  (\bm{m}\K\m^T\K)-s\frac{\beta(\beta-1)}{4\Delta}\Tr (\Q\K\Q^T\K)\nonumber\\
    &-s\frac{\beta^2(\tau-1)}{4\Delta}\Tr ((\q+\eps)\K(\q+\eps)^T\K)\nonumber\\
    &-\frac{(\beta\tau)^2(\frac{s}{\tau}(\frac{s}{\tau}-1))}{4\Delta}\Tr (\q\K\q^T\K),
\end{align}
while the channel term $I(\hat{\T})$ reads
\begin{align}
    &e^{\sum\limits_{a'a\alpha\leq c'c\gamma}\mathrm{tr}\left(\hat{\T}_{a'a\alpha,c'c\gamma}
    \bm{x}^{c'c\gamma}
    \bm{x}^{a'a\alpha T}\right)}\nonumber\\
    &=\mathbb{E}^{\mathcal{N}(0,1)}_{\bm{\xi}}\prod\limits_{a'=1} \mathbb{E}^{\mathcal{N}(0,1)}_{\bm{\lambda}_{a'}}\prod\limits_{a'a} \mathbb{E}^{\mathcal{N}(0,1)}_{\bm{\eta}_{a'a}}e^{(\bm{x}^{0})^{T}\bm{\hat{m}}\sum\limits_{a'a\alpha\ne0}\bm{x^{a'a\alpha}}
    }\nonumber\\
    &\times e^{-\frac{1}{2}\sum\limits_{a'a\alpha\ne0}(\bm{x^{a'a\alpha}})^{T}
    \bm{\hat{Q}}\bm{x^{a'a\alpha}}
    +\sum\limits_{a'a}\bm{\eta}_{a'a}^T(\hat{\Q}-\hat{\q}-\hat{\eps})^{\frac{1}{2}}\sum\limits_\alpha \x^{a'a\alpha}}\nonumber\\
    &\times e^{+\sum\limits_{a'}\bm{\lambda}^T_{a'}\hat{\eps}^{\frac{1}{2}}
    \sum\limits_{a\alpha}\bm{x^{a'a\alpha}}+\bm{\xi}^{T}\bm{\hat{q}}^{\frac{1}{2}}\sum\limits_{a\alpha\ne0}x^{a\alpha}}.
\end{align}
Here, $\bm{\xi}$, $\{\bm{\lambda_{a'}}\}_{a'=1}^{\frac{s}{\tau}}$ and $\{\bm{\eta_{a'a}}\}_{a',a\in\llbracket 1,\frac{s}{\tau}\rrbracket\times\llbracket 1,\tau\rrbracket}$ are again Hubbard Stratonovitch fields following a normal distribution.
Rearranging the terms, we get
\begin{align}
    I(\hat{\T})=&\frac{s}{\tau}\mathbb{E}^{\mathcal{N}(0,1)}_{\bm{\xi}}\mathbb{E}^{P_X}_{\x^0}\ln \Bigg\{
    \mathbb{E}^{\mathcal{N}(0,1)}_{\bm{\lambda}}\Bigg[
    \mathbb{E}^{\mathcal{N}(0,1)}_{\bm{\eta}}e^{\beta h(\x^0)}\nonumber\\&
    +\mathbb{E}^{\mathcal{N}(0,1)}_{\bm{\eta}}\Big(\int d\x P_X(\x)e^{h(\x)}\Big)^\beta
    \Bigg]^\tau
    \Bigg\}
\end{align}
with
\begin{align}
    h(\x)\overset{\rm{def}}{=}&-\frac{1}{2\Delta}\x^T\Q\x
    +\frac{1}{\Delta^{\frac{1}{2}}}\Big(\frac{1}{\Delta^{\frac{1}{2}}}\x^{0T}\m+\bm{\eta}^T(\Q-\q-\eps)^{\frac{1}{2}}\nonumber\\
    &+\bm{\lambda}^T\eps^{\frac{1}{2}}+\bm{\xi}^T\q^{\frac{1}{2}}\Big)\x.
\end{align}
The 1RSB free entropy then reads
\begin{widetext}
\begin{align}
\label{eq:1rsb}
    &\Phi(\beta,\phi)^{1\rm{RSB}}=-\frac{\beta}{2\Delta}\Tr  (\bm{m}\K\m^T\K)-\frac{\beta(\beta-1)}{4\Delta}\Tr (\Q\K\Q^T\K)-\frac{\beta^2(\tau-1)}{4\Delta}\Tr ((\q+\eps)\K(\q+\eps)^T\K)+\frac{\beta^2\tau}{4\Delta}\Tr (\q\K\q^T\K)\nonumber\\
    &
    +\frac{1}{\tau}\mathbb{E}^{\mathcal{N}(0,1)}_{\bm{\xi}}\mathbb{E}^{P_X}_{\x^0}\ln \Bigg\{
    \mathbb{E}^{\mathcal{N}(0,1)}_{\bm{\lambda}}\Bigg[
    \mathbb{E}^{\mathcal{N}(0,1)}_{\bm{\eta}}e^{\beta h(\x^0)}
    +\mathbb{E}^{\mathcal{N}(0,1)}_{\bm{\eta}}\Big(\int d\x P_X(\x)e^{h(\x)}\Big)^\beta
    \Bigg]^\tau
    \Bigg\}.
\end{align}
\end{widetext}
\normalsize

\subsection{Stability analysis}

\begin{figure}
    \centering
    \includegraphics[scale=.6]{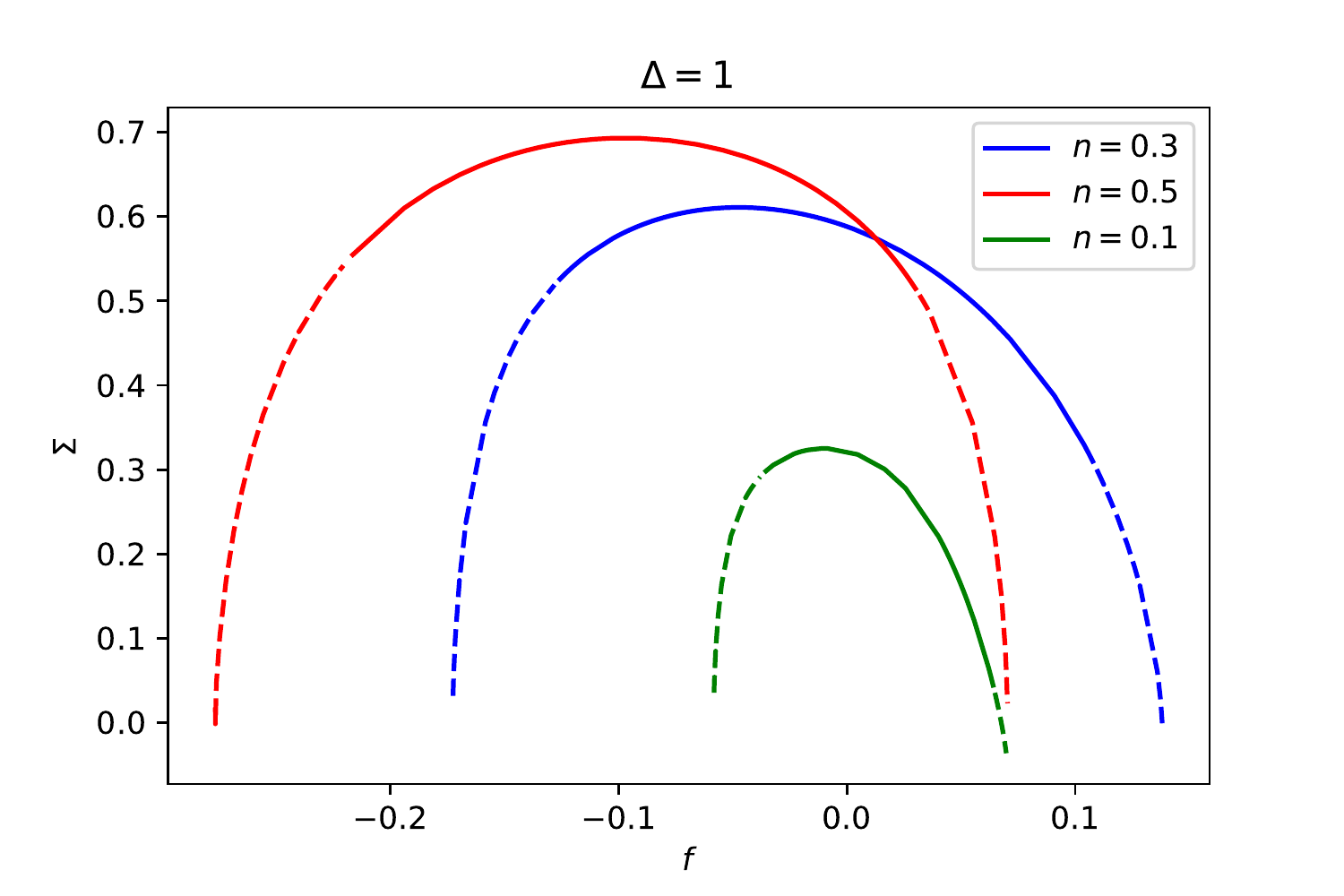}
    \caption{Subset-entropy curves for an SBM \eqref{eq:SBM_prior}\eqref{eq:SBM_posterior} with $\Delta=1$, see also Fig.~\ref{fig:Sigma_curves}. Regions where the RS fixed point \eqref{eq:ansatz_begin}-\eqref{eq:ansatz_end} is stable (as evaluated using the stability condition \eqref{eq:stability}) are represented in solid lines; unstable regions are dashed.}
    \label{fig:stability}
\end{figure}

The 1RSB free entropy \eqref{eq:1rsb} can be expanded in orders of $\eps^{\frac{1}{2}}$,
\begin{align}
&\Phi(\beta,\phi)^{1\rm{RSB}}=\nonumber\\
&\Phi^{(0)}(\beta,\phi)+\Tr(\eps^{\frac{1}{2}}\Phi^{(\frac{1}{2})}(\beta,\phi))+\Tr(\eps\Phi^{(1)}(\beta,\phi))+o(\eps),
\end{align}
with the zeroth order term corresponding to the RS free entropy introduced in equation \eqref{eq:final_entro}. The computation of order $\eps^{\frac{1}{2}}$ terms and above requires the computation of matrix derivative and is in general hard to evaluate without further assumption on the structure of $\eps$. We now restrict ourselves to the special case analyzed in this work, namely to rank $r=1$ where all order parameters are scalars. 

In the rank 1 case, introducing the shorthand
\begin{align}
    \mathcal{Z}(z)\overset{\rm{def}}{=}e^{\phi-\frac{\beta}{2\Delta}Q(x^0)^2+\beta z\times x^0}+\left(\int dx P_X(x)e^{-\frac{1}{2\Delta}Qx^2+z\times x}
    \right)^\beta,
\end{align}
with
\begin{align}
    z\overset{\rm{def}}{=}\frac{1}{\Delta^{\frac{1}{2}}}\left(\frac{1}{\Delta^{\frac{1}{2}}}x^0m+\eta(Q-q)^{\frac{1}{2}}+\xi q^{\frac{1}{2}}\right),
\end{align}
the expansion can be carried out up to order $\epsilon$. It can be shown that $\Phi^{(\frac{1}{2})}(\beta,\phi)=0$ while
\begin{align}
   &\Phi^{(1)}(\beta,\phi)\nonumber\\&=\frac{\tau-1}{\Delta}\left[-\beta^2q+\mathbb{E}^{\mathcal{N}(0,1)}_{\bm{\xi}}\mathbb{E}^{P_X}_{\x^0}\left(\frac{\mathbb{E}^{\mathcal{N}(0,1)}_\eta \Z'(z)}{\mathbb{E}^{\mathcal{N}(0,1)}_\eta \Z(z)}\right)^2\right].
\end{align}
This $o(\epsilon)$ correction yields another term in the saddle point condition $\partial_q \Phi^{\mathrm{1RSB}}\overset{!}{=}0$. The zeroth order part of this condition is identical to the RS saddle-point for $q$, and is automatically satisfied, while requiring the $\mathcal{O}(\epsilon)$ part of the gradient to be vanishing implies
\begin{align}
    \epsilon=\frac{1}{\beta^2}\mathbb{E}^{\mathcal{N}(0,1)}_{\bm{\xi}}\mathbb{E}^{P_X}_{\x^0}\partial_q\left(\frac{\mathbb{E}^{\mathcal{N}(0,1)}_\eta \Z'(z)}{\mathbb{E}^{\mathcal{N}(0,1)}_\eta \Z(z)}\right)^2\epsilon.
\end{align}
If the term multiplying $\epsilon$ in the RHS has absolute value greater than $1$, iterating the saddle point equation causes the perturbation to grow, signalling a lack of stability of the RS fixed point. Expanding the derivative the stability condition yields
\begin{widetext}
\begin{equation}
\label{eq:stability}
\frac{1}{\Delta\beta^2}\Bigg|\mathbb{E}^{\mathcal{N}(0,1)}_{\bm{\xi}}\mathbb{E}^{P_X}_{\x^0}\left(\frac{\mathbb{E}^{\mathcal{N}(0,1)}_\eta \Z''(z)}{\mathbb{E}^{\mathcal{N}(0,1)}_\eta \Z(z)}\right)^2-4\mathbb{E}^{\mathcal{N}(0,1)}_{\bm{\xi}}\mathbb{E}^{P_X}_{\x^0}\frac{(\mathbb{E}^{\mathcal{N}(0,1)}_\eta \Z'(z))^2\mathbb{E}^{\mathcal{N}(0,1)}_\eta \Z''(z)}{(\mathbb{E}^{\mathcal{N}(0,1)}_\eta \Z(z))^3}+3\mathbb{E}^{\mathcal{N}(0,1)}_{\bm{\xi}}\mathbb{E}^{P_X}_{\x^0}\left(\frac{\mathbb{E}^{\mathcal{N}(0,1)}_\eta \Z'(z)}{\mathbb{E}^{\mathcal{N}(0,1)}_\eta \Z(z)}\right)^4\Bigg|<1.
\end{equation}
\end{widetext}

\paragraph{Stability condition for the SBM with two balanced communities}

We specialize the stability condition \eqref{eq:stability} to the two-balanced communities SBM setup defined by \eqref{eq:SBM_posterior} and \eqref{eq:SBM_prior}. We employ the  notation \eqref{eq:shorthand_SBM_1}-\eqref{eq:shorthand_SBM_2}, with the addition of
\begin{equation}
    V_x=\frac{1}{Z_x}(\rho x_+^2 e^{h_+}+(1-\rho) x_-^2 e^{h_-})-f_x^2,
\end{equation}
which can be seen as the variance of the measure $P_X(\cdot)e^{h(\cdot)}$, as would naturally appear in an Approximate Message Passing setting for example \cite{Thibault_Constrained,Thibault_MMSE,review,CS}. We give here the expressions of $\mathbb{E}^{\mathcal{N}(0,1)}_\eta\Z,\mathbb{E}^{\mathcal{N}(0,1)}_\eta\Z',\mathbb{E}^{\mathcal{N}(0,1)}_\eta\Z''$ that are involved in \eqref{eq:stability}:
\begin{align}
    &\mathbb{E}^{\mathcal{N}(0,1)}_\eta\Z=e^{\phi+\beta h_0}+\mathbb{E}^{\mathcal{N}(0,1)}_\lambda Z_x^\beta,\\
    &\mathbb{E}^{\mathcal{N}(0,1)}_\eta\Z'=\beta x^0 e^{\phi+\beta h_0}+\beta \mathbb{E}^{\mathcal{N}(0,1)}_\lambda Z_x^\beta f_x,
\end{align}
and
\begin{align}
    \mathbb{E}^{\mathcal{N}(0,1)}_\eta\Z''=&
    \beta^2 (x^0)^2e^{\phi+\beta h_0}\nonumber\\
    &+\beta\mathbb{E}^{\mathcal{N}(0,1)}_\lambda \Big(
    (\beta-1)Z_x^{\beta}f_x^2+Z_x^{\beta}(V_x+f_x^2)
    \Big).
 \end{align}
These expression can be readily plugged back into \eqref{eq:stability} to yield the stability condition for the SBM with two balanced communities.\\

\paragraph{Results} The evaluation of the stability condition \eqref{eq:stability} for the particular setting of the SBM with two balanced communities is shown in Fig.~\ref{fig:stability}. As expected, the assumed replica symmetry is validated in proximity of the maximum of the curves, corresponding to the typical case (with maximal subset-entropy) obtained with uniform random sampling of the labelled nodes. On the other hand, the results of the large deviation computation are not found to be exact when the described events become very rare and the correlations grow too much. Note that the curves become unstable before reaching negative values of the subset-entropy, as expected.
While a complete analysis of the fixed point of the RSB free energy \eqref{eq:1rsb} (beyond infinitesimal perturbations from the RS fixed point) would be needed in this case, it is numerically more involved and we leave it for future work. Note, however, that RSB corrections are typically small in magnitude \cite{RSB} and the numerical simulations seem to be in good agreement with the obtained theoretical predictions, despite the instability of the ansatz.

\begin{algorithm}[H]
\caption{MCf, MCm}
\label{alg:MCMC}
\begin{algorithmic}
   \STATE {\bfseries Input:} budget $n$, adjacency matrix $Y$, ground truth labels $\X^0$, criterion$=m$ (MCm) or $f$ (MCf) evaluated using TAP, convergence criterion $\epsilon$, max number of iteration $M$.
   \STATE Initialize $\SSel$ randomly by subsampling $nN$ nodes uniformly at random.
   \STATE Initialize number of iterations $t=0$, difference $\delta=0$
    \WHILE{$\delta>\epsilon$ and $t<M$}
    \STATE $t=t+1$
    \STATE sample $i\in \SSel$ at random and $j\in\llbracket1,N\rrbracket\smallsetminus\sub$ at random\\
    \STATE name $\SSel'=\SSel\smallsetminus\{i\}\cup\{j\}$
    \STATE compute criterion($\SSel$') using TAP (Algorithm \ref{alg:TAP})
    \IF{criterion($\SSel'$)$>$criterion($\SSel$)}
    \STATE $\SSel\leftarrow \SSel'$
    \ENDIF
    \ENDWHILE
\end{algorithmic}
\end{algorithm}

\begin{algorithm}[H]
\caption{semi-supervised TAP}
\label{alg:TAP}
\begin{algorithmic}
   \STATE {\bfseries Input:} disclosed subset $\SSel$, adjacency matrix $Y$ convergence criterion $\epsilon$, max number of iteration $M$.
   \STATE Initialize $\forall i, \hat{\x}_i^{t=0}\in\mathbb{R}^{r}\leftarrow 10^{-3}\mathcal{N}(0,\mathbbm{1}_r)$
   \STATE Initialize $\forall i, \bm{V}_i^{t=0}\in\mathbb{R}^{r}\leftarrow0$
   \STATE Initialize $ \bm{U}^{t=0}\in\mathbb{R}^{r\times r}\leftarrow0$
   \STATE Initialize $\forall i, \bm{\sigma}_i^{t=0}\in\mathbb{R}^{r\times r}\leftarrow0$
    \WHILE{$\frac{1}{N}\sum\limits_{i}||\hat{\x}^{t}_i-\hat{\x}^{t-1}_i||>\epsilon$ and $t<M$}
    \STATE$\bm{\sigma}_i^{t+1}=\partial_{\bm{V}}f_i(\bm{U}^{t},\bm{V}_i^{t})$
    \STATE$\hat{\x}_i^{t+1}=f_i(\bm{U}^{t},\bm{V}_i^{t})$
   \STATE$
    \bm{U}^{t+1}=\frac{1}{N\Delta}\sum\limits_k (\K\hat{\x}_k^{t+1})(\K\hat{\x}_k^{t+1})^T$
    \STATE$\bm{V}_i^{t+1}=\frac{1}{\sqrt{N}}\sum\limits_kS_{ik}\K\hat{\x}_k^{t+1}-\frac{1}{N\Delta}\sum\limits_k\K\bm{\sigma}_k^{t+1}\K\hat{\x}_i^{t}$
   
    \STATE where
    \STATE $f_i(\bm{U},\bm{V}_i)=\frac{\int d\x  P^\SSel_{X|i}(\x) \x e^{-\frac{1}{2}\x^T\bm{U}\x+\x^T\bm{V}_i}}{\int d\x  P^\SSel_{X|i}(\x) e^{-\frac{1}{2}\x^T\bm{U}\x+\x^T\bm{V}_i}}$
    \STATE $P^\SSel_{X|i}(\x)=(\mathbbm{1}_{i\notin\SSel}P_X(\x)+\mathbbm{1}_{i\in\SSel}\delta(\x-\x^0_i))$
    \STATE $S_{ij}=\partial_w g(Y_{ij}|w)|_{w=0}$
    \ENDWHILE
\RETURN
\STATE The free energy 
\begin{align}
    f&=-\sum\limits_i\mathrm{ln}\int d\x  P^\SSel_{X|i}(\x) e^{-\frac{1}{2}\x^T\bm{U}\x+\x^T\bm{V}_i}\nonumber\\
    &+\frac{1}{2N}\Bigg[
    \frac{1}{\sqrt{N}}\sum\limits_{i,k}S_{ik}\hat{\x}_i^T
    \hat{\x}_k-2\mathrm{tr}\left(\bm{U}\sum\limits_k\bm{\sigma}_k\right)\nonumber\\
    &-\frac{1}{2\Delta N}\sum\limits_{i,k}\mathrm{tr}\left(\K\hat{\x}_i\hat{\x}_i^T\K\hat{\x}_k\hat{\x}_k^T\right)
    \Bigg]
\end{align}
\STATE The magnetization $m=\frac{1}{N}\sum\limits_{i=1}^N \x_i^0\hat{\x}_i^T$
\end{algorithmic}
\end{algorithm}

\section{Consequences for active learning} \label{sec:consequences}

\begin{figure}
    \centering
    \includegraphics[scale=.6]{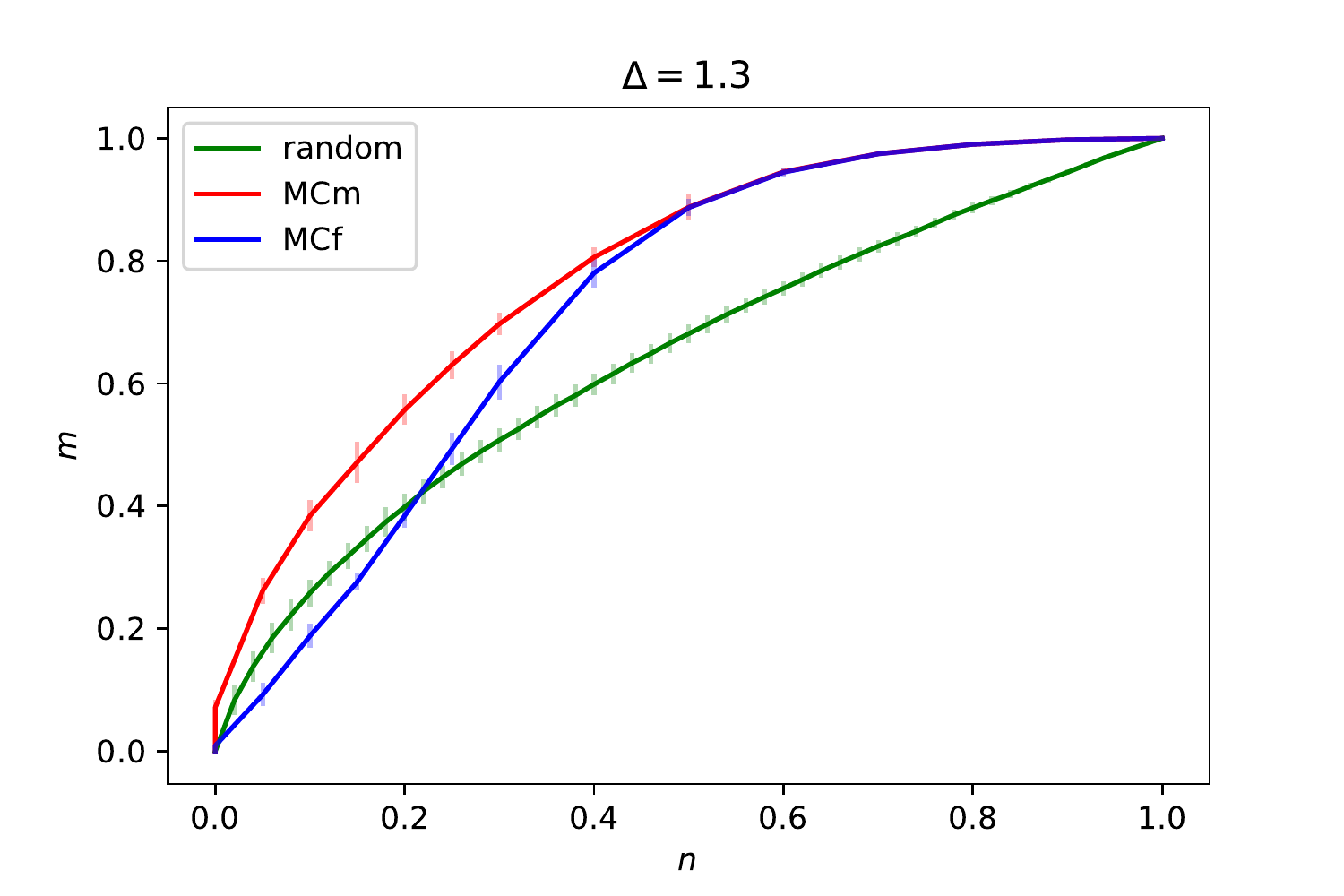}\\
    \includegraphics[scale=.6]{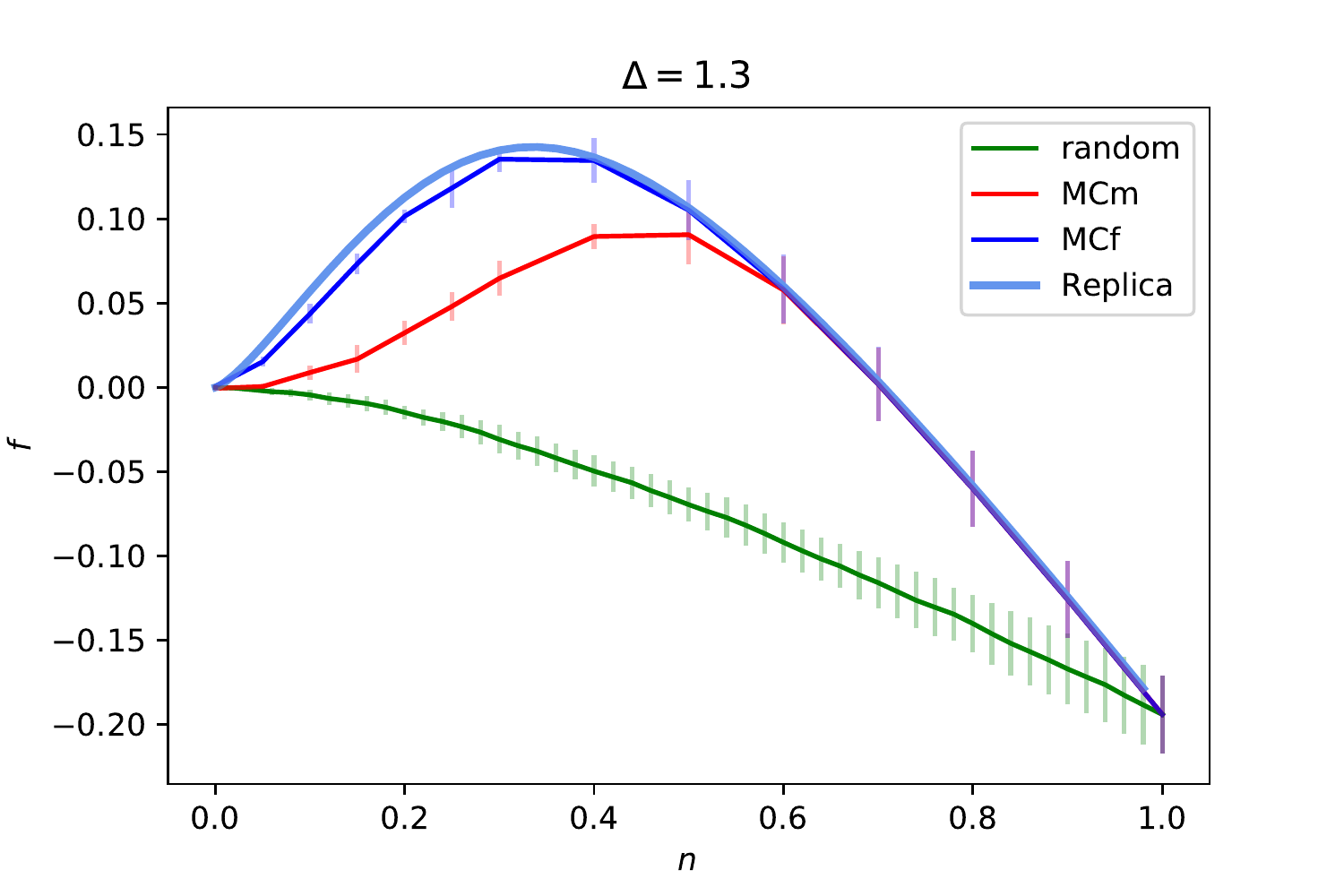}
    \caption{SBM with two balanced communities \eqref{eq:SBM_posterior}\eqref{eq:SBM_prior} with $N=10^3$ nodes and low SNR ($\Delta=1.3$) (Upper) magnetizations $m$ achieved by the MCm and MCf algorithms, benchmarked against random subset selection, for various budgets $n$. All points are averaged over $40$ SBM instances. Maximizing $m$ and $f$ cease to be equivalent below $n^\star\approx 0.6$. For $n\lessapprox 0.3$ maximizing the free energy $f$ leads to sub-random performances in terms of magnetization, signalling the onset of the relationship between $f$ and $m$ becoming non-monotonous, see Figs.~\ref{fig:Sigma_curves} and \ref{fig:critical} (Lower) free energies $f$ achieved by the MCm and MCf algorithms (Algorithm \ref{alg:MCMC}), benchmarked against random subset selection, and the highest achievable free energy predicted by the large deviation \eqref{eq:LD_principle} (light blue) $\mathrm{inf}\{f|\Sigma(n,f)>0\}$, see also Fig.~\ref{fig:Sigma_curves}. MCf consistently returns the theoretical highest free energy. }
    \label{fig:heuristics_Del_1.3}
\end{figure}

\begin{figure}
    \centering
    \includegraphics[scale=.6]{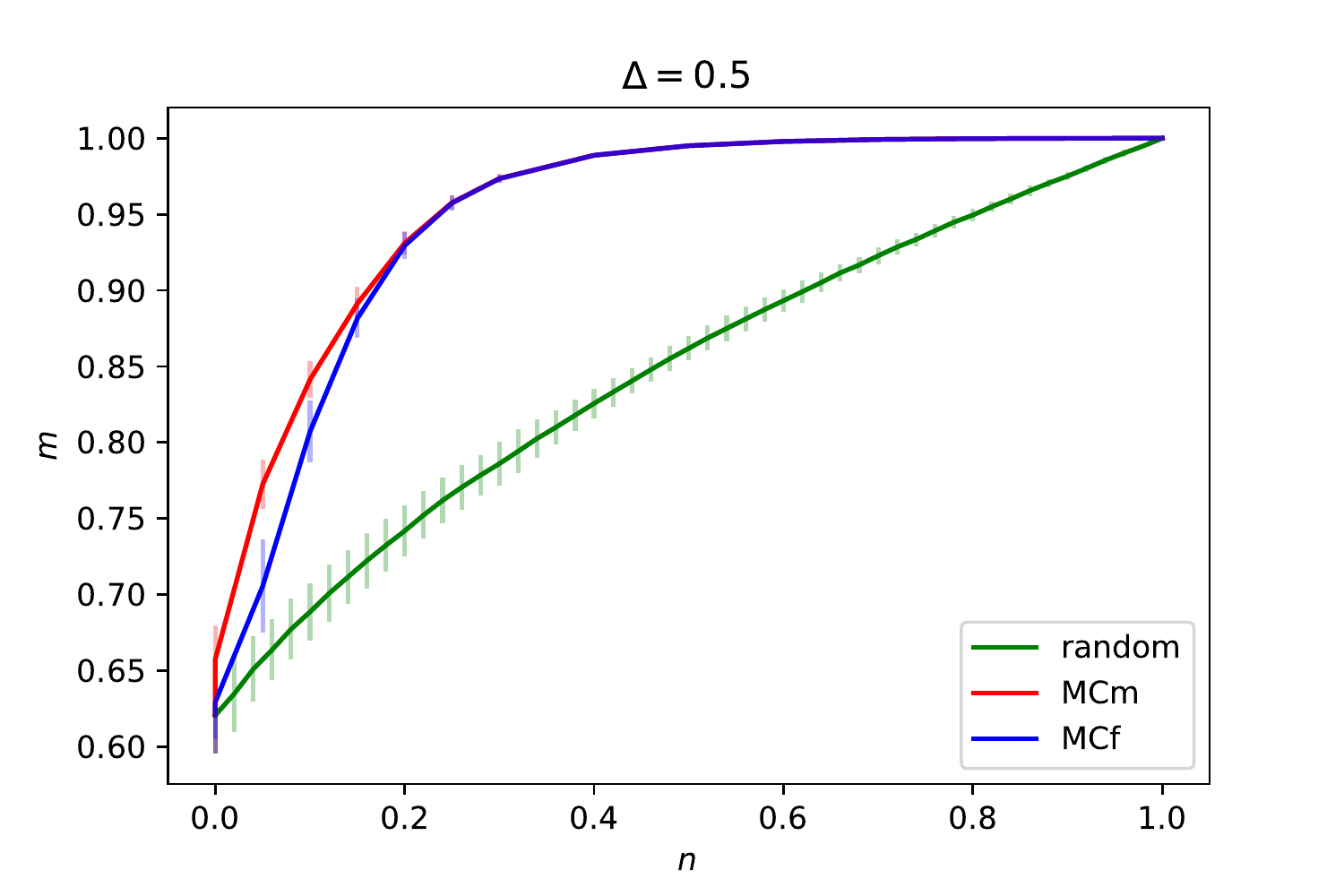}\\
    \includegraphics[scale=.6]{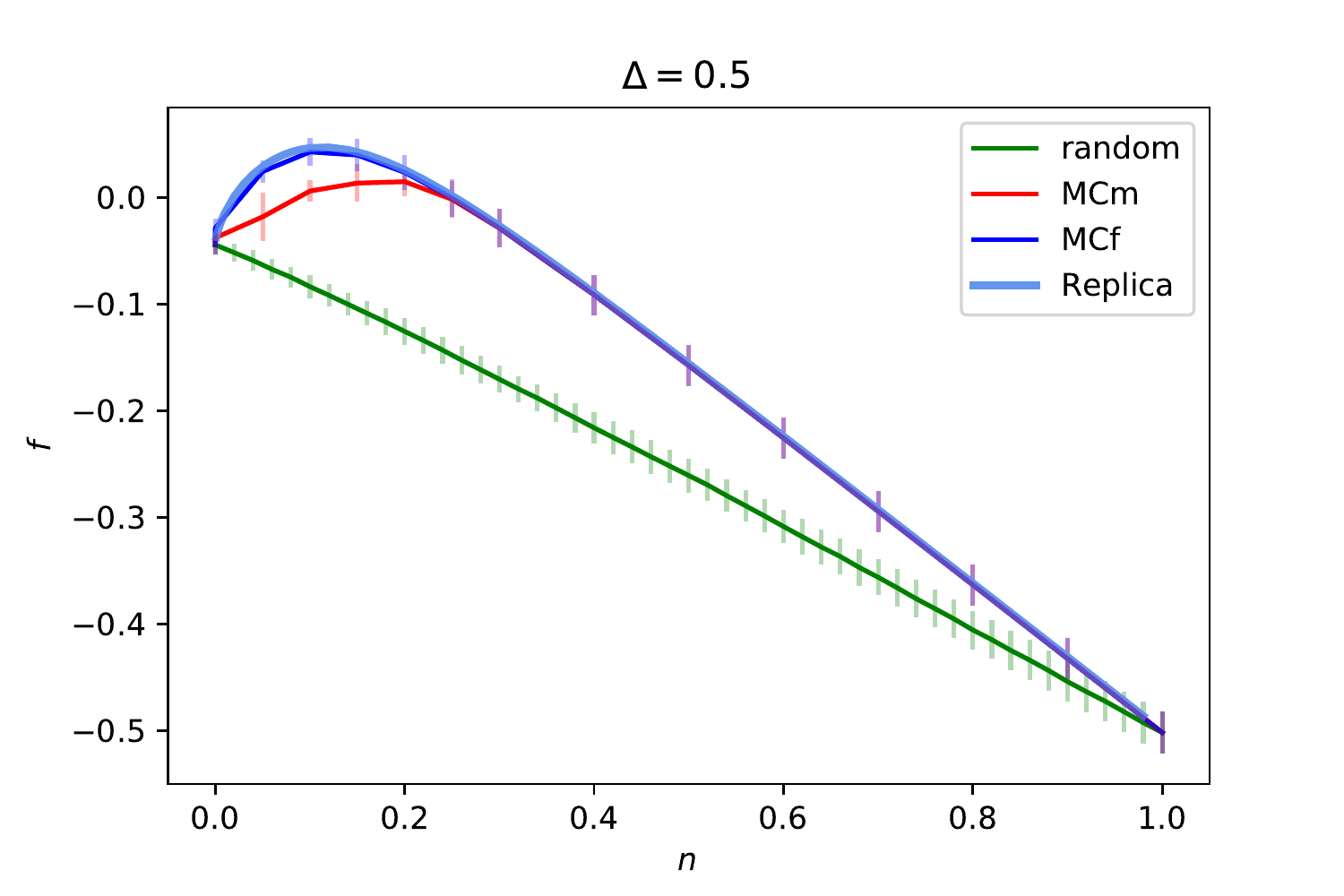}
    \caption{SBM with two balanced communities \eqref{eq:SBM_posterior}\eqref{eq:SBM_prior} with $N=10^3$ nodes and high SNR ($\Delta=0.5$) (Upper) magnetizations $m$ achieved by the MCm and MCf algorithms, benchmarked against random subset selection, for various budgets $n$. All points are averaged over $40$ SBM instances. Maximizing $m$ and $f$ cease to be equivalent below $n^\star\approx 0.3$. (Lower) free energies $f$ achieved by the MCm and MCf algorithms (Algorithm \ref{alg:MCMC}), benchmarked against random subset selection, and the highest achievable free energy predicted by the large deviation \eqref{eq:LD_principle} (light blue) $\mathrm{inf}\{f|\Sigma(n,f)>0\}$, see also Fig.~\ref{fig:Sigma_curves}. MCf consistently returns the theoretical highest free energy. }
    \label{fig:heuristics_Del_0.5}
\end{figure}

It has been shown in \cite{Cui2020LargeDF} for perceptron learning that computing the maximal free energy from large deviations could provide a valuable benchmark for assessing the effectiveness and the limitations of \textit{active learning} strategies. We discuss in this section how much this relationship carries over (or differs) in the case of active community detection in the SBM. In the active community inference problem, the statistician is allowed to choose for which subset of nodes (of size corresponding to a fraction $n$) the true communities are revealed, and leverage on this additional knowledge to better infer the ground truth community assignment. The goal is therefore to find an algorithmic strategy for determining a set $\mathcal{S}$ with cardinal $|S|=nN$ of individuals, from which the maximal amount of information can be extracted once their true communities is revealed.\\

A first remark is that it would of course seem more natural to look directly at the large deviation properties for the reconstruction error \eqref{eq:reconstruction}. Carrying a large-deviation analysis \eqref{eq:LD_principle} for $\mathrm{tr}\bm{m}$ is, however, a challenging and still open problem. In \cite{Cui2020LargeDF}, the free energy \eqref{eq:free_energy_matrixfacto} associated to an inference problem was proposed in the context of perceptron learning as a possible proxy for the magnetization, with the advantage of having a computable large deviation. It is therefore tempting to adopt the free energy, $f$, as a measure of informativeness for the disclosed subset $\mathcal{S}$. However, as evidenced in Fig.~\ref{fig:Sigma_curves} and discussed in \ref{sec:results}, for sufficiently noisy problems (sufficiently large $\Delta$) and low budgets $n$, the free energy ceases to be a monotonously increasing function of the magnetization $m$. As a consequence, finding the subset $\mathcal{S}$ leading to largest free energy is \textit{not always} tantamount to finding the subset leading to maximal magnetization. \\

\begin{figure}
    \centering
    \includegraphics[scale=0.6]{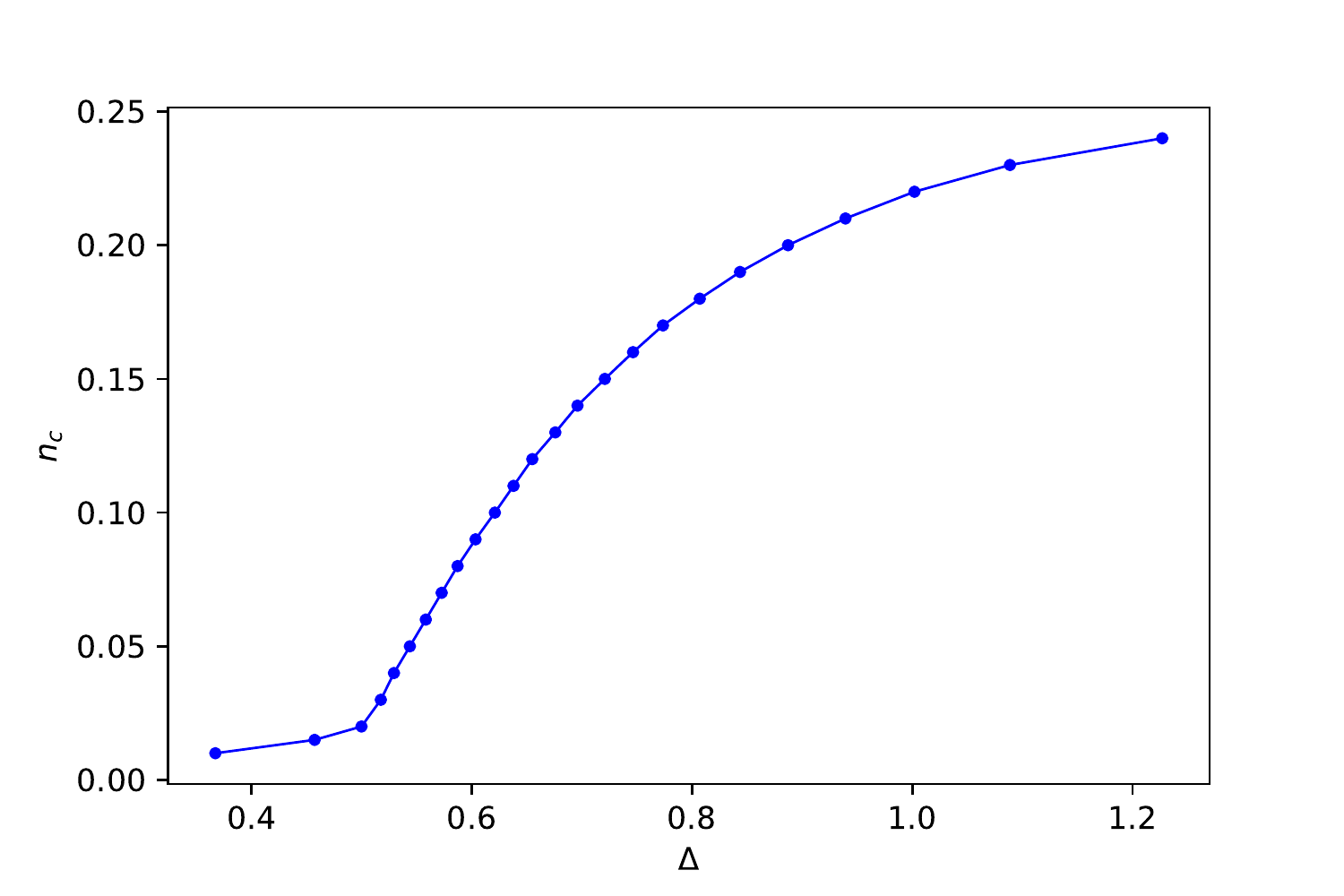}
    \caption{Critical value of the budget $n$ at which non-monotonous relationship between free-energy and reconstruction overlap can be observed, as a function of the noise level $\Delta$. }
    \label{fig:critical}
\end{figure}

\begin{figure}
    \centering
    \includegraphics[scale=0.6]{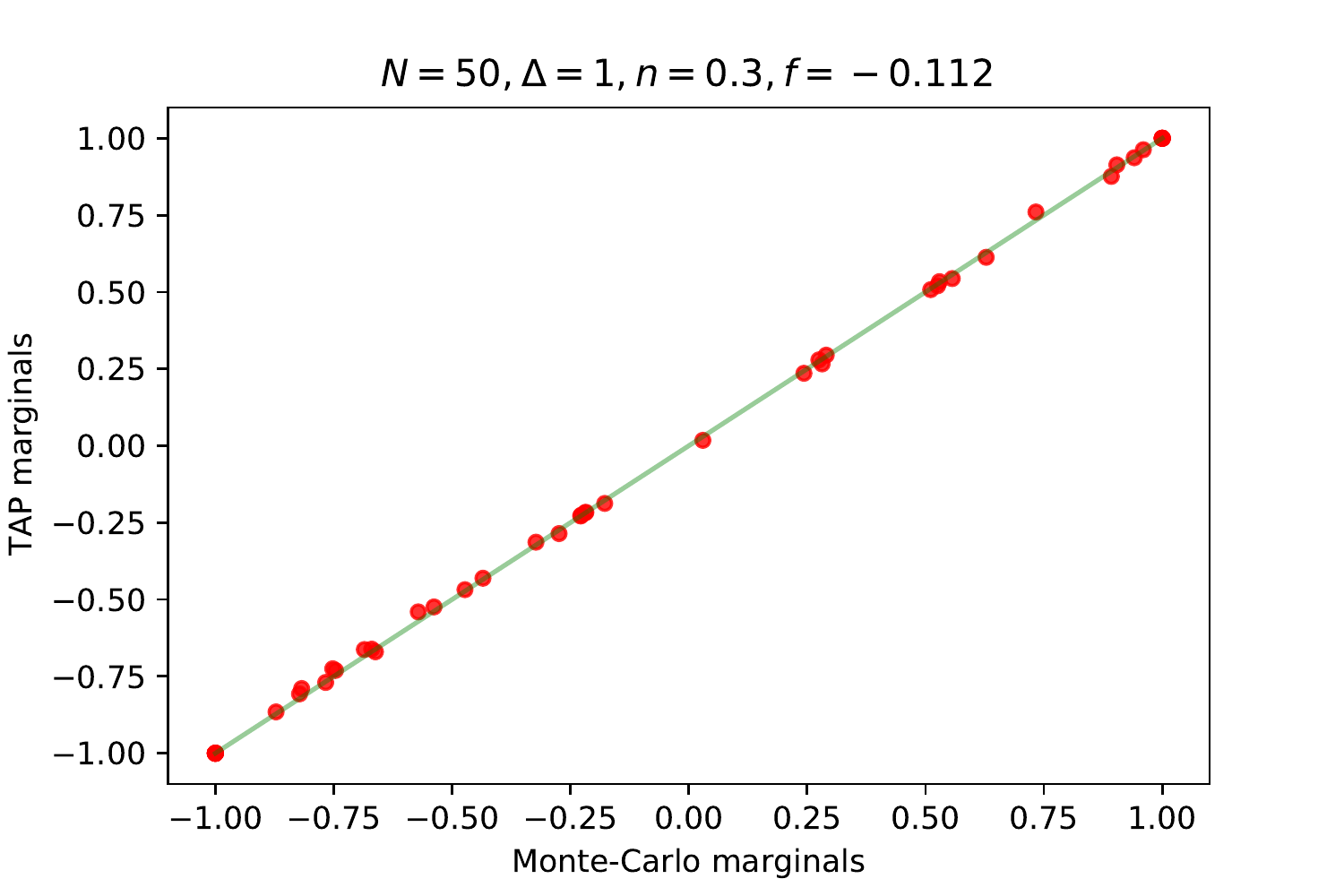}
    \caption{Scatter plot of the local magnetization for each node $x_i$ of a SBM \eqref{eq:SBM_prior}\eqref{eq:SBM_posterior} on $N=50$ nodes as estimated per TAP, versus its estimation using a standard Metropolis-Hastings Monte Carlo sampling ($10^7$ iterations, and a $50$ steps are given to let the system thermalize between every two sampled configurations). A selection $\SSel$ corresponding to a large deviation of the free energy (Fig.~\ref{fig:Sigma_curves}) $f\approx-0.0112$ is disclosed. Despite the correlations introduced by the latter, the marginals given by TAP are consistent with the ones obtained with Monte Carlo. }
    \label{fig:MCMC}
\end{figure}

In Fig.~\ref{fig:heuristics_Del_1.3} and \ref{fig:heuristics_Del_0.5}, the subsets leading to maximal $f$ (resp. $m$) were numerically searched on an instance of a SBM community detection problem using greedy Monte Carlo schemes, MCf (resp. MCm), see Algorithm \ref{alg:MCMC}. For each value of the budget $n$, these schemes were initialized on a random subset of size $nN$. At each iteration, a replacement of a selected node by an unselected node was proposed: the resulting change in $f$ (or $m$) was evaluated by estimating the posterior marginals using the Thouless-Anderson-Palmer (TAP) message-passing inference algorithm \cite{TAP,Thibault_Constrained,Thibault_MMSE} (Algorithm \ref{alg:TAP}), and accepted if leading to an increase of the objective. Though there exists a priori no guarantee that TAP correctly computes the posterior marginals when some nodes are disclosed, we found that the marginals returned by TAP stood in very good agreement with those estimated using a standard (but slower) Monte-Carlo sampling, see Fig. \ref{fig:MCMC}.  The performance of the Monte Carlo schemes MCf and MCm were contrasted with the one of random subset selection, and benchmarked against the maximal attainable $f$ as extracted from the theoretical entropy curve $\Sigma(n,f)$, in Fig.~\ref{fig:Sigma_curves}. The MCf maximization scheme consistently returns for each budget $n$ the theoretical highest achievable free energy predicted by our large deviation analysis, $f_{\mathrm{max}}^n=\mathrm{argmax}\{f|\Sigma(n,f)>0\}$ (Fig.~\ref{fig:Sigma_curves}), see the agreement of the dark blue and light blue curves in Figs.~\ref{fig:heuristics_Del_0.5}, \ref{fig:heuristics_Del_1.3}. For the considered values of $\Delta=0.5,~1.3$, there exists a budget $n^\star$ below which a gap between MCm and MCf appears: in particular, MCf does not achieve the maximal magnetization, i.e. maximizing the free energy $f$ ceases to be equivalent to maximizing the magnetization. Therefore, if an active learning algorithm was capable of indirectly maximizing $f$, the obtained performance would be sub-optimal. Moreover, note that the MCf algorithm analyzed in this section exploits direct knowledge of all the ground-truth labels, and cannot be applied in a realistic active learning setting. The value of the budget $n^\star$ below which this discrepancy becomes observable decreases for lower $\Delta$ (less noisy settings). The critical values for the budgets, extrapolated from the large deviation, are shown in Fig.~\ref{fig:critical}. Note that, for the most noise settings, for instance $\Delta=1.3$ in Fig.~\ref{fig:heuristics_Del_1.3}, selecting the subset $\mathcal{S}$ with largest $f$ even yields sub-random performance in terms of magnetization for small budgets, signalling that in this range of $n$ the magnetization and the free energy cease to covary.
By using a straightforward generalization of Algorithms \ref{alg:MCMC},\ref{alg:TAP}, we checked that an analogous phenomenon can also be traced in sparse SBMs \cite{Decelle2011AsymptoticAO} when the noise level (controlled by the difference between the inter and intra community connectivities) approaches the associated critical value.

\vspace{-2mm}
\section{Conclusions}

In this work, we considered the semi-supervised learning setting in the dense SBM. Through a statistical physics analysis based on the replica method, we derived an analytic description of the large deviation properties of the model under the different possible choices of the subsets of labelled nodes. 
The results of our computation show that there exists rare subsets with higher (lower) information content with respect to typical samples, and allowed for a characterization of the relationship between information content and rareness. Moreover, we found that lower SNR levels in the inference problem can lead to a non-monotonous relationship between the volume of the log-posterior measure and the reconstruction accuracy associated to different subset selections. 
In order to validate our theoretical predictions, we identified the regions of stability for the replica symmetric ansatz (assumed in the calculations) and we displayed different comparisons of our analytic results with numerical simulations in finite graph instances.
Finally, we discussed the connection of the presented large deviation analysis with the active learning problem in community reconstruction. We thus identified the region of parameters where our analysis can yield a good approximation for the optimal performance achievable with any active selection algorithm.

Possible future research directions involve looking at the algorithmic implications of our analysis, especially in the case of sparse networks. Moreover, the non-trivial finding of the impact of the SNR in selection processes could be further explored in different settings (e.g., supervised learning problems \cite{Cui2020LargeDF}) where active learning strategies are commonly employed.

\begin{acknowledgments}
We acknowledge funding from the ERC under the European Union’s Horizon 2020 Research and Innovation Program Grant Agreement 714608-SMiLe.
\end{acknowledgments}

\nocite{*}

\bibliography{biblio.bib}

\newpage
\appendix

\end{document}